\documentclass[twocolumn,twocolappendix]{aastex631}

\usepackage{graphicx}
\usepackage{xcolor,graphicx,xspace,color,longtable}
\usepackage{amssymb,amsmath,shadow,bezier,curves,rotating}
\usepackage{graphicx,chngcntr}
\usepackage{multirow} 
\usepackage{enumitem}
\graphicspath{ {./images/}}

\newcommand {\h}    {$h^{-1}\,\mathrm{Mpc}$}

\newcommand {\hm}   {$h^{-1} \  M_{\odot}$}

\newcommand {\m}    {$M_{\odot}$}

\begin{document}

\title[Anisotropy Velocity Profile in Uchuu Simulations] {Quantifying the Velocity Anisotropy Profile of Galaxy Clusters Using the Uchuu Cosmological Simulation}

\author[0000-0003-3595-7147]{Mohamed H. Abdullah}
\affiliation{Department of Physics, University of California Merced, 5200 North Lake Road, Merced, CA 95343, USA}
\affiliation{Department of Astronomy, National Research Institute of Astronomy and Geophysics, Cairo, 11421, Egypt}

\author[0009-0003-7202-4159]{Raouf H. Mabrouk}
\affiliation{Department of Astronomy, National Research Institute of Astronomy and Geophysics, Cairo, 11421, Egypt}
\affiliation{Department of Astronomy, Space Science, and Meteorology, Faculty of Science, Cairo University, Giza
11326, Egypt}

\author[0000-0002-5316-9171]{Tomoaki Ishiyama}
\affiliation{Digital transformation enhancement council, Chiba University, 1-33, Yayoi-cho, Inage-ku, Chiba, 263-8522, Japan}

\author[0000-0002-6572-7089]{Gillian Wilson}
\affiliation{Department of Physics, University of California Merced, 5200 North Lake Road, Merced, CA 95343, USA}

\author{Magdy Y. Amin}
\affiliation{Department of Astronomy, Space Science, and Meteorology, Faculty of Science, Cairo University, Giza
11326, Egypt}

\author{Elamira Hend Khattab}
\affiliation{Department of Astronomy, Space Science, and Meteorology, Faculty of Science, Cairo University, Giza
11326, Egypt}

\author{H. I. Abdel Rahman}
\affiliation{Department of Astronomy, National Research Institute of Astronomy and Geophysics, Cairo, 11421, Egypt}

\begin{abstract}
Galaxy clusters are powerful laboratories for studying both cosmic structure formation and galaxy evolution. We present a comprehensive analysis of the velocity anisotropy profile, $\beta(r)$, in galaxy clusters using the Uchuu-UniverseMachine mock galaxy catalog, which combines the large-volume Uchuu $N$-body simulation with the UniverseMachine galaxy formation model. Focusing on clusters with $\log{\mathrm{M_{200}}} \geq 13.9\ [h^{-1} \mathrm{M_\odot}]$ up to redshift $z=1.5$, we investigate the behavior of $\beta(r)$ as a function of cluster-centric radius, mass, and redshift. We find that $\beta(r)$ exhibits a universal shape: it rises from isotropic values near the cluster core, peaks at $\sim1.7\mathrm{R_{200}}$, declines around $3.4\mathrm{R_{200}}$ due to orbital mixing, and increases again in the outskirts due to the dominance of first-infalling galaxies. Our results show that more massive clusters have higher radial anisotropy and larger peak $\beta$ values. Moreover, $\beta(r)$ evolves with redshift, with high-redshift clusters displaying more radially dominated orbits and enhanced infall motions. We further derive redshift-dependent power-law scaling relations between $\mathrm{M_{200}}$ and key physical radii—hydrostatic ($\mathrm{R}_{\mathrm{hs}}$), infall ($\mathrm{R}_{\mathrm{inf}}$), and turnaround ($\mathrm{R}_{\mathrm{ta}}$).These findings offer a robust theoretical framework for interpreting the dynamical properties of observed galaxy clusters, and provide key insights into the evolution of their dynamical state over cosmic time
\end{abstract}


\section{Introduction}

Galaxy clusters are the largest gravitationally bound structures in the universe, making them essential to our understanding of both astrophysics and cosmology \citep{Voit05,Kravtsov12}. From a cosmological perspective, galaxy clusters provide unique insights into the universe evolution. Their abundance and distribution over cosmic time are directly tied to cosmological parameters, offering valuable constraints on the nature of dark energy and the growth of structure \citep{Vikhlinin09,Allen11,Abdullah23,Ishiyama2025}. From an astrophysical standpoint, the dense environments of galaxy clusters shape the evolution of galaxies within them. Physical processes such as dynamical friction, ram pressure stripping, galaxy harassment, and strangulation play crucial roles in altering galaxy morphology, star formation activity, and gas content \citep{Tyler13,Ebeling14,Boselli14,Tagliaferro21}.

A key dynamical property of galaxy clusters is the velocity anisotropy profile, $\beta(r)$, which provides information about the orbital structure of galaxies within the cluster potential \citep{Wojtak09,Biviano04}. The $\beta(r)$ profile serves as a tracer of the cluster relaxation state, revealing whether it has undergone recent mergers or is dynamically evolving \citep{Hou09}. Additionally, $\beta(r)$ is critical for improving the accuracy of dynamical mass estimates, which are essential for cosmological applications \citep{The86,Wojtak10}. Furthermore, $\beta(r)$ reflects the dark matter potential well, as it depends on the orbital distribution of galaxies responding to the cluster’s overall gravitational field \citep{Host09}.

The study of velocity anisotropy in galaxy clusters has a long history. Early theoretical work by \citet{Merritt85} explored the orbital dynamics of galaxies, laying the foundation for understanding anisotropy profiles. Observational studies followed, with measurements of velocity dispersions in galaxy clusters \citep{Carlberg97,Biviano04}. These studies identified a general trend of increasing radial anisotropy with cluster-centric radius. However, observational limitations, particularly projection effects in redshift space, made it difficult to disentangle the radial and tangential velocity components \citep{Praton94,Abdullah13}.

Advances in cosmological simulations have opened new opportunities for studying velocity anisotropy. High-resolution simulations provide the means to analyze $\beta(r)$ in unprecedented detail across a wide range of masses and redshifts. 
For instance, \citet{Wetzel11} showed that recently accreted satellite galaxies tend to follow more radial orbits, particularly in the outer regions of clusters, highlighting the strong connection between a cluster’s accretion history and the orbital structure of its galaxy population. \citet{Mamon13} showed that dynamically relaxed clusters tend to exhibit smoother and more stable velocity anisotropy profiles compared to merging systems, which display more variable and disturbed $\beta(r)$ trends due to recent dynamical activity. The role of baryonic physics in shaping the velocity structure of clusters has been highlighted in hydrodynamical simulations \citep{Munari13}, where processes such as gas cooling, star formation, and feedback modify the kinematics of galaxies, particularly in the core regions. These effects can lead to lower velocity dispersion and potentially reduced anisotropy in cluster centers compared to dark matter-only simulations. More recently, \citet{Lotz19} examined $\beta(r)$ in hydrodynamical simulations from the Magneticum project \citep{Hirschmann14,Dolag15}, and found that baryonic physics, cluster mass, dynamical state, and redshift all play significant roles in shaping the form and normalization of $\beta(r)$, particularly in the central regions of clusters.
Despite these advances, discrepancies persist between observed and simulated velocity anisotropy profiles, especially in the cores of clusters \citep{Oman13}. These inconsistencies highlight gaps in our understanding of cluster dynamics and motivate further studies to resolve these issues.

In this work, we investigate the velocity anisotropy profile, $\beta(r)$, using the Uchuu-UniverseMachine \citep[Uchuu-UM:][]{Aung23} mock galaxy catalog. This catalog combines the UniverseMachine algorithm \citep{Behroozi19} with the Uchuu cosmological simulation \citep{Ishiyama21}, a large-volume, high-resolution $N$-body simulation spanning $8 \, \mathrm{h^{-3}} \, \mathrm{Gpc}^3$ spatial volume. Its extensive volume and high mass resolution ($3.27 \times 10^8 \, h^{-1} \, \mathrm{M_\odot}$ particle mass) make it ideally suited for studying galaxy clusters across a broad range of masses and redshifts \citep{Aung23}. Our analysis focuses on the dependence of $\beta(r)$ on cluster mass and redshift, leveraging the strengths of the Uchuu-UM simulation to explore these relationships in detail. Simulated data enable us to investigate kinematic trends that are difficult to discern in observational datasets due to projection effects and sample limitations.

In addition to characterizing $\beta(r)$, we also derive redshift-dependent power-law scaling relations between $\mathrm{M_{200}}$ and key physical radii—hydrostatic ($\mathrm{R}_{\mathrm{hs}}$), infall ($\mathrm{R}_{\mathrm{inf}}$), and turnaround ($\mathrm{R}_{\mathrm{ta}}$), which we define in Section \ref{sec:veldist}. These relations provide a theoretical bridge between observable quantities and the underlying dynamics of clusters, and are essential for refining mass estimates and understanding cluster evolution in a cosmological context. Ultimately, this work contributes to the broader effort of using galaxy clusters as precise tools for cosmological research.

This paper is organized as follows. Section~\ref{sec:Theory} provides a theoretical overview of velocity anisotropy and introduces the Uchuu-UM simulation. Section~\ref{sec:results} presents our results, highlighting the variation of $\beta(r)$ with cluster mass and redshift, and introducing the redshift-dependent scaling relations. Finally, Section~\ref{sec:conc} summarizes our findings and outlines future directions. Throughout this paper, we assume a $\Lambda$CDM cosmology with $\Omega_\mathrm{m} = 1 - \Omega_\Lambda$ and $h_0 = 100 \, h~\mathrm{km\,s^{-1}\,Mpc^{-1}}$. All logarithms are base-10, denoted by $\log$.

\section{Theoretical Framework and Simulated Data} \label{sec:Theory}

In this section, we outline the theoretical basis for studying the velocity anisotropy profile in galaxy clusters and describe the simulated data used in our analysis.

\subsection{Velocity Anisotropy in Galaxy Clusters}\label{sec:beta}

The velocity anisotropy parameter, $\beta(r)$, provides a quantitative measure of how galaxy orbits in a cluster deviate from isotropy. It is a crucial tool for understanding the orbital structure and dynamical state of galaxy clusters. Mathematically, $\beta(r)$ is defined as:
\begin{equation}\label{eq:beta}
\beta(r) = 1 - \frac{\sigma_{\theta}^2 + \sigma_{\phi}^2}{2 \sigma_r^2} \equiv 1 - \frac{\sigma_t^2}{\sigma_r^2},
\end{equation}
\noindent where $\sigma_r$ is the radial velocity dispersion, $\sigma_{\theta}$ and $\sigma_{\phi}$ are the tangential velocity dispersions ($\sigma_{t}$) in the polar and azimuthal directions, respectively, which are typically equal, $\sigma_{\theta}$ = $\sigma_{\phi}$. The anisotropy parameter $\beta(r)$ ranges from $-\infty$ (for purely tangential motion) to $1$ (for purely radial motion), with $\beta = 0$ corresponding to isotropic orbits. If $\beta > 0$, the system is biased towards radial motion, while if $\beta < 0$, it is biased towards tangential motion \citep{Binney87}.

Typically, $\beta(r)$ trends reveal valuable information about the internal dynamics of clusters. Near the cluster center, $\beta(r)$ is often close to zero, indicating nearly isotropic orbits \citep{Biviano13}. Moving outward, $\beta(r)$ tends to increase, suggesting a preference for radial orbits in the outer regions \citep{Wojtak09}. Beyond the virial radius, some studies suggest that $\beta(r)$ plateaus or even decreases slightly \citep{Lemze12}. 

\begin{table}
    \centering
    \caption{Summary of the number of galaxy clusters used in this study. Columns show the redshift, the mass range, the average cluster mass in that bin, and the number of clusters ($\mathrm{N_c}$).}
    \begin{tabular}{c c c c c}
    \hline
    $z$ & \multicolumn{2}{c}{mass interval} & $\log \langle \mathrm{M_{200}} \rangle$ & $\mathrm{N_c}$ \\
    & \multicolumn{2}{c}{$[h^{-1}~\mathrm{M_\odot}]$} & {$[h^{-1}~\mathrm{M_\odot}]$} & \\
    \hline
    \multicolumn{5}{c}{Number of clusters for four mass bins at $z=0$} \\
    \hline
    0.00  & \multicolumn{2}{c}{$13.9 \leq \log \mathrm{M_{200}} < 14.2$} & 14.02 & 197,318 \\
    0.00  & \multicolumn{2}{c}{$14.2 \leq \log \mathrm{M_{200}} < 14.5$} & 14.31 & 73,422  \\
    0.00  & \multicolumn{2}{c}{$14.5 \leq \log \mathrm{M_{200}} < 14.9$} & 14.62 & 24,163  \\
    0.00  & \multicolumn{2}{c}{$14.9 \leq \log \mathrm{M_{200}} $} & 14.99 & 2,454   \\
    \hline
    \multicolumn{5}{c}{Number of clusters at four redshifts} \\
    \hline
    0.00  & \multicolumn{2}{c}{$13.9 \leq \log \mathrm{M_{200}} < 15.7$} & 14.10 & 297,357 \\
    0.49  & \multicolumn{2}{c}{$13.9 \leq \log \mathrm{M_{200}} < 15.3$} & 14.06 & 139,165 \\
    1.03  & \multicolumn{2}{c}{$13.9 \leq \log \mathrm{M_{200}} < 15.1$} & 14.02 & 37,443  \\
    1.54  & \multicolumn{2}{c}{$13.9 \leq \log \mathrm{M_{200}} < 14.8$} & 13.99 & 7,284   \\
     \hline
    \end{tabular}
\label{tab:cluster_summary}
\end{table}

The velocity anisotropy profile of galaxies in clusters is often modeled using simple parametric functions that capture the transition from isotropy in the central regions to radially biased orbits in the outskirts. One such model is the \citet{Tiret07} anisotropy profile, given by  
\begin{equation}
\beta(r) = \beta_{\infty} \mathrm{\frac{r}{r + r_{\beta}}},
\end{equation}
where $\beta_{\infty}$ represents the asymptotic anisotropy at large radii, and $\mathrm{r}_{\beta}$ is a scale radius that determines the transition between the inner isotropic and  outer radially anisotropic regions.
At small radii ($\mathrm{r} \ll \mathrm{r}_{\beta}$), the velocity distribution remains nearly isotropic ($\beta \approx 0$), while at large radii ($\mathrm{r} \gg \mathrm{r}_{\beta}$), the profile asymptotically approaches $\beta_{\infty}$. This model provides a simple yet effective description of the velocity anisotropy transition in galaxy clusters and has been widely used in numerical simulations and observational studies to characterize the orbital distribution of galaxies in different environments \citep{Mamon13}. Other models, such as those by \citet{Carollo95}, \citet{Baes07}, and \citet{Mamon13}, account for additional complexities. However, the exact shape of $\beta(r)$ can vary significantly across clusters and is influenced by factors such as the cluster dynamical state, mass, and formation history \citep{Mamon10}.

The connection between velocity anisotropy and the mass distribution of galaxy clusters is encapsulated in the Jeans equation. For a spherically symmetric system, this equation takes the form:
\begin{equation} \label{Jeans}
M_J(r) = -\frac{r\sigma_r^2}{G} \left[\frac{d\ln{\rho}}{d\ln{r}} + \frac{d\ln{\sigma_r^2}}{d\ln{r}} + 2\beta(r)\right],
\end{equation}
where $M_J(r)$ is the mass enclosed within radius $r$, $\rho(r)$ is the galaxy number density profile, and $G$ is the gravitational constant \citep{Binney08,Tagliaferro21}. This equation highlights the critical role of $\beta(r)$ in connecting the observed velocity dispersion to the underlying mass distribution, making it a fundamental component in dynamical mass estimates for galaxy clusters \citep{Mamon13}.

The mass distribution of galaxy clusters is often well-described by the Navarro-Frenk-White (NFW) profile \citep{NFW96,NFW97}:
\begin{equation}
\rho(r) = \frac{\rho_0}{\mathrm{\frac{r}{r_s} \left(1 + \frac{r}{r_s}\right)^2}},
\end{equation}
where $\rho_0$ is a characteristic density, and $\mathrm{r_s}$ is the scale radius. This model is widely used for dark matter-dominated systems and provides a robust framework for interpreting velocity anisotropy profiles.

\begin{figure*}
    \centering    
    \includegraphics[width=1\linewidth]{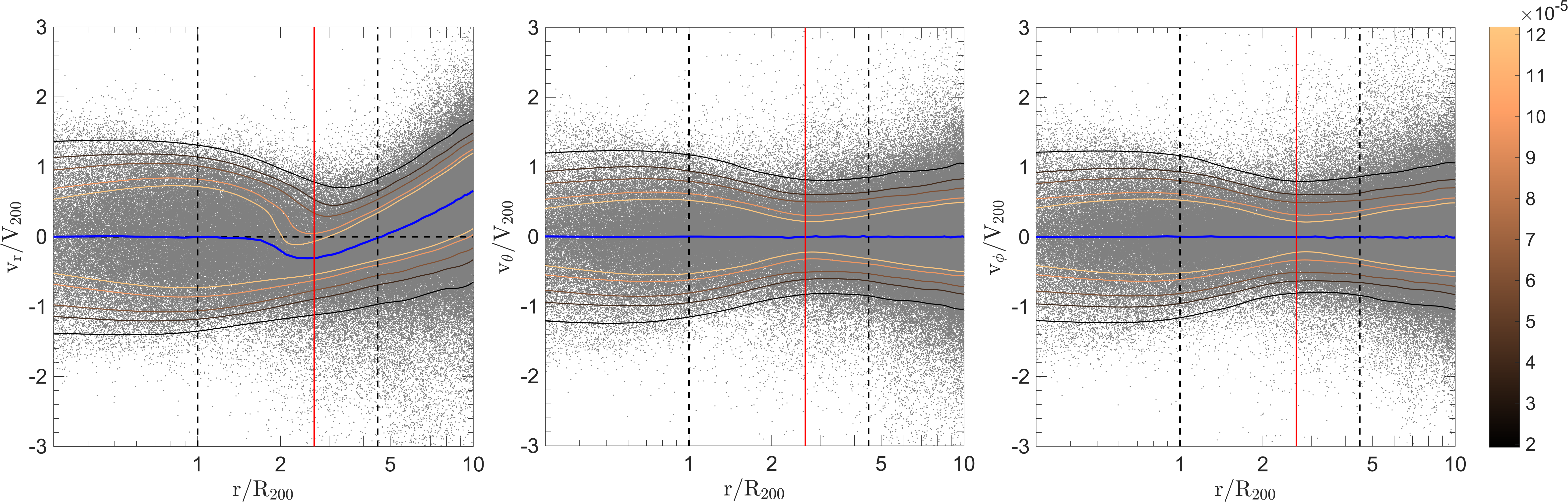}
    \caption{The distribution of normalized velocities $(\mathrm{v}/\mathrm{V_{200}})$ for galaxies in 1000 stacked galaxy clusters, with masses in the range $13.9 \leq \log{\mathrm{M}_{200}} \leq 15.3~[h^{-1}~\mathrm{M_{\odot}}]$ and galaxies with stellar masses $\log{\mathrm{M_s}} \geq 9.36~[h^{-2}~\mathrm{M_{\odot}}]$ ($\sim 0.1$ stellar mass of the Milky Way Galaxy) at redshift $z = 0$, is plotted as a function of normalized radius $(\mathrm{r/R_{200}})$. The three panels show the radial velocity component $(\mathrm{v_r})$ (left), and the two tangential velocity components, $(\mathrm{v_\theta})$ (middle) and $(\mathrm{v_\phi})$ (right). The solid blue line in each panel represents the median velocity $\langle \mathrm{v/V_{200}} \rangle$ as a function of distance from the cluster center.
    The two vertical dashed black lines mark the hydrostatic radius ($\mathrm{R_{hs}}$) and the turnaround radius ($\mathrm{R_{ta}}$), both defined where $\mathrm{v/V_{200}} \approx 0$, while the red solid line indicates the infall radius ($\mathrm{R_{inf}}$), corresponding to the minimum radial velocity. Note that in this figure we assume $\mathrm{R_{hs}} = \mathrm{R_{200}}$; however, in Section~\ref{sec:scalingrelation}, we present a method for determining $\mathrm{R_{hs}}$ directly from the velocity profile.
    The horizontal dashed black line in the left panel shows the zero velocity line. Each panel includes 2D adaptive kernel smoothed (2DAKM) contours at levels of 60, 70, 80, 90, and 95 percent, indicating the concentration of galaxies in each velocity component as a function of radius.}
    \label{fig:meanvel}
\end{figure*}
\subsection{The Uchuu-UM Galaxy Simulation}
Our study uses data from the Uchuu-UM mock galaxy catalog \citep{Aung23}, derived from the Uchuu cosmological simulation \citep{Ishiyama21}. Uchuu is part of a suite of large, high-resolution $N$-body simulations designed to model the evolution of dark matter structures in a $\Lambda$CDM cosmology consistent with $Planck$ 2015 parameters \citep{Planck15}: $\Omega_0 = 0.3089$, $\Omega_b = 0.0486$, $\lambda_0 = 0.6911$, $h = 0.6774$, $\mathrm{n}_s = 0.9667$, and $\sigma_8 = 0.8159$. The simulation box spans $2000 \, (h^{-1} \, \mathrm{Mpc})$ on each side, with a particle mass resolution of $3.27 \times 10^8 \, (h^{-1} \, \mathrm{M_\odot})$ and a gravitational softening length of $4.27 \, (h^{-1} \, \mathrm{kpc})$.

The Uchuu simulation utilized the \textsc{greem} code \citep{Ishiyama09,Ishiyama12} for the $N$-body calculations, with halos and subhalos identified using the \textsc{rockstar} algorithm \citep{Behroozi13a} and merger trees constructed via \textsc{consistent trees} \citep{Behroozi13b}. We use $\mathrm{M}_{200}$ as the halo mass, which is defined as the mass within an overdensity of 200 times the critical density of the Universe. We define $\mathrm{R}_{200}$ as the radius within which the cluster is approximately in hydrostatic equilibrium where the overdensity is 200 times the critical density of the Universe. We also define $\mathrm{V}_{200} = \sqrt{G \mathrm{M_{200}/R_{200}}}$ as the circular velocity of the cluster, where G is the gravitational constant.

The Uchuu-UM Galaxy catalog was created using the UniverseMachine model \citep{Behroozi19}, which assigns galaxy properties to dark matter halos based on their assembly histories. The model parametrizes star formation rates as functions of halo mass, growth history, and redshift. Stellar masses are computed by integrating these star formation rates over time, accounting for stellar mass loss. The UniverseMachine parameters were optimized using a Markov Chain Monte Carlo algorithm to match a variety of observational datasets, including stellar mass functions, cosmic star formation rates, and UV luminosity functions, across a wide range of redshifts ($0 < z < 10$). 

Compared to earlier simulations such as MultiDark \citep{Klypin16}, Uchuu-UM provides a significantly larger statistical sample, with eight times the volume and five times the mass resolution. These advantages make it ideal for studying the velocity anisotropy profiles of galaxy clusters across diverse masses and redshifts. Halo and subhalo catalogs, along with merger trees, are publicly available through the Skies \& Universes website.\footnote{\url{http://www.skiesanduniverses.org/Simulations/Uchuu/}}. In this paper, we focus on clusters with masses $\log{\mathrm{M}_{200}} \geq 13.9~[h^{-1}~\mathrm{M_{\odot}}]$. Table~\ref{tab:cluster_summary} summarizes the number of clusters used in our analysis, divided by mass bins at $z = 0$ and across different redshifts.

\section{Results and Discussion} \label{sec:results}
In this section, we present the results of the velocity anisotropy profile, $\beta(r)$. We first examine the velocity distribution of galaxies in the cluster field. We then discuss the overall behavior of the $\beta(r)$ profile for the entire sample at redshift $z=0$, followed by a detailed analysis of how $\beta(r)$ varies as a function of both cluster mass and redshift.

\subsection{Velocity Distribution of galaxies in the cluster field} \label{sec:veldist}
Before discussing the velocity anisotropy profile, we first examine the velocity distributions of galaxies in 1,000 stacked galaxy clusters, randomly selected from the full sample, at redshift $z=0$. Figure \ref{fig:meanvel} illustrates the distribution of normalized velocities, $\mathrm{(v/V_{200}})$ as a function of normalized radius, $(\mathrm{r/R_{200}})$. The figure shows the distribution of  galaxies for 1000 stacked clusters, with masses in the range $13.9 \leq \log{\mathrm{M}_{200}}$ [\hm] $\leq 15.3$ and galaxies with stellar masses $\log{\mathrm{M_s}}~[h^{-2}~\mathrm{M_{\odot}}] \geq 9.36$ ($\sim 0.1$ stellar mass of the Milky Way Galaxy) at redshift $z = 0$. 
The three panels represent the radial velocity component, $\mathrm{v_r}$ (left), and the two tangential velocity components, $\mathrm{v_\theta}$ (middle) and $\mathrm{v_\phi}$ (right). The contours in each panel represent the 2D adaptive kernel smoothed (2DAKM) distribution of galaxy velocities, with contour levels at 60, 70, 80, 90, and 95 percent, highlighting the concentration of galaxies at different velocities and radii. These components are analyzed to investigate the kinematics of galaxies in different regions of the clusters and how these velocities change with radius.

As presented in the left panel of Figure \ref{fig:meanvel}, the ensemble phase-space density can be categorized into three main regions: (i) an inner hydrostatic region, where $\langle \mathrm{v_r} \rangle=0$, indicating a relatively relaxed state; (ii) 
an infall region with $\langle \mathrm{v_r} \rangle <0$, containing two opposing streams of material—one moving inward toward the halo (which may include both first-time infallers and material undergoing a second or subsequent infall), and another moving outward after reaching its closest approach;
and (iii) an outflow region, where $\langle \mathrm{v_r} \rangle>0$, dominated by the perturbed Hubble flow. Most material in regions (i) and (ii) remains gravitationally bound to the halo, whereas almost all material in region (iii) is unbound. The characteristic scale that separates relaxed and infall regions is referred to as the hydrostatic radius, $\mathrm{R_{hs}}$ (marked by the first vertical dashed black line, see \citealp{Busha05,Cuesta08}), where $0.75~\mathrm{R}_{200}\lesssim \mathrm{R_{hs}} \lesssim 1.25~\mathrm{R}_{200}$. The characteristic scale that separates the infall and outflow regions is called the turnaround radius $\mathrm{R_{ta}}$ (marked by the second vertical dashed black line, see \citealp{Gunn72,Abdullah13}), where $\mathrm{R_{ta}} \simeq 4.5~\mathrm{R}_{200}$.

\begin{figure*}
    \centering
    \includegraphics[width=1\linewidth]{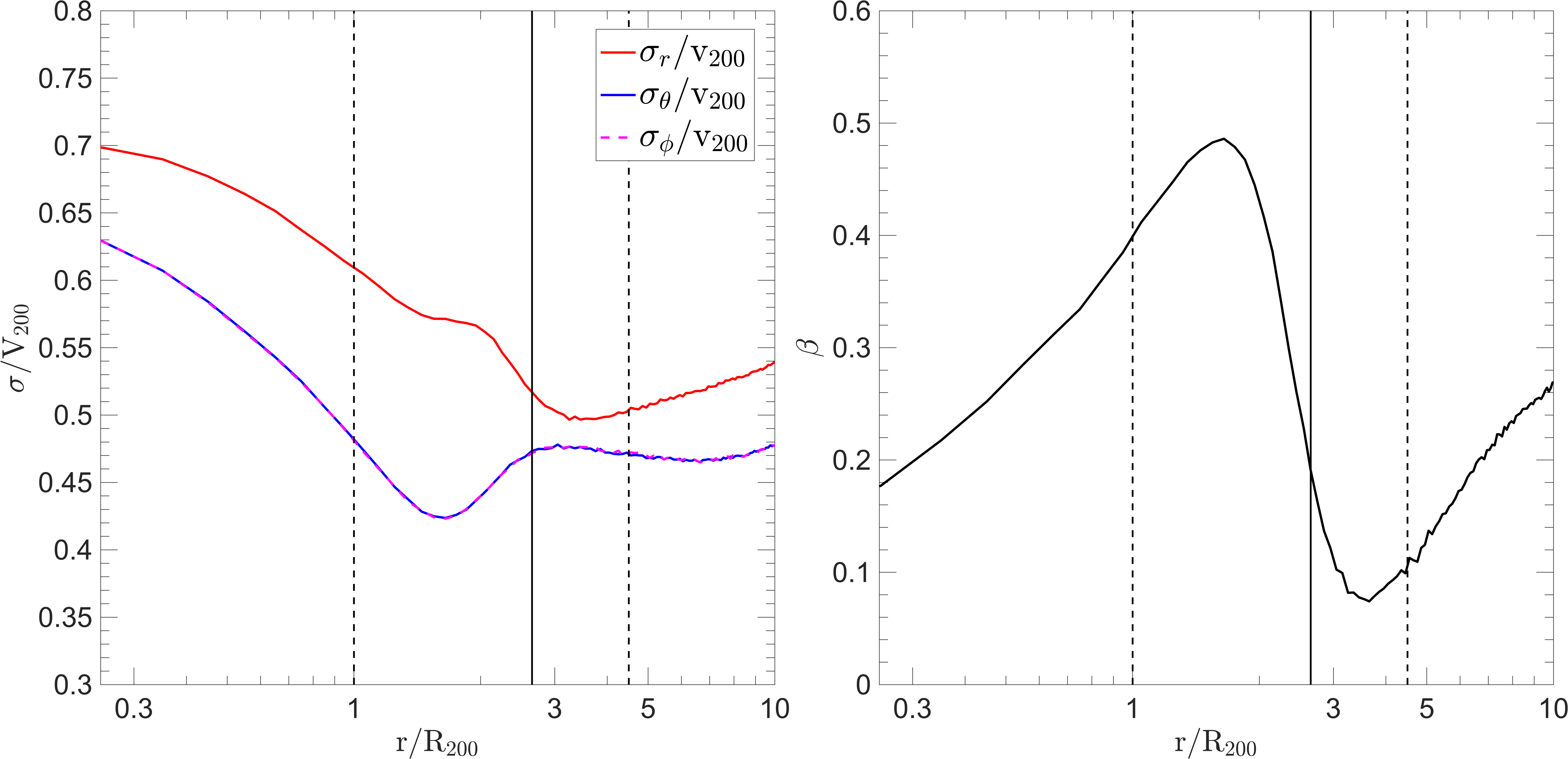}
    \caption{Velocity anisotropy analysis for galaxy clusters with masses in the range $13.9 \leq \log{\mathrm{M}_{200}} \leq 15.3~[h^{-1}~\mathrm{M_\odot}]$ and galaxies with stellar masses $\log{\mathrm{M_s}} \geq 9.36~[h^{-2}~\mathrm{M_\odot}]$ at redshift $z = 0$. The left panel shows the radial ($\sigma_r$), polar ($\sigma_\theta$), and azimuthal ($\sigma_\phi$) velocity dispersions, each normalized by the virial velocity $\mathrm{V}_{200}$, as a function of normalized cluster-centric radius ($\mathrm{R}/\mathrm{R}_{200}$). The right panel displays the total velocity anisotropy profile, $\beta(R)$, averaged over all clusters. The two vertical dashed black lines mark the hydrostatic radius ($\mathrm{R_{hs}}$) and the turnaround radius ($\mathrm{R_{ta}}$), both defined where $\mathrm{v}/\mathrm{v_{200}} \approx 0$. The solid black vertical line indicates the infall radius ($\mathrm{R_{inf}}$), corresponding to the location of minimum radial velocity.
}
    \label{fig:betaall}
\end{figure*}

From the cluster core ($\mathrm{r} = 0$) to the hydrostatic radius ($\mathrm{R_{hs}}$), galaxy clusters are generally in a virialized state. In this region, galaxies move in random orbits under the influence of the cluster's gravitational potential, resulting in a median radial velocity close to zero. This reflects a state of virial equilibrium, where inward and outward motions of galaxies are balanced in a dynamically relaxed system. Beyond the hydrostatic radius, in the infall region between $\mathrm{R_{hs}}$ and $\mathrm{R_{ta}}$, the median radial velocity starts to deviate from zero, becoming increasingly negative and reaching a minimum value around $2.65~\mathrm{R_{200}}$. This minimum marks the radius of the maximum infall velocity, which we designate as the infall radius, $\mathrm{R_{inf}}$ \citep{Wetzel15} (denoted by the vertical solid red line). At $\mathrm{R_{inf}}$, the infall velocity $\mathrm{v_{inf}}$ reaches its maximum value. This velocity is expressed as: $\mathrm{v_{inf}(r)}=\mathrm{H_{0}r + v_{pec}(r)}$, where $\mathrm{H_{0}}$ is the Hubble constant and $\mathrm{v_{pec}(r)}$ is the peculiar velocity \citep{Abdullah11}. 
This infall radius arises because $\mathrm{v_r}$ is calculated as the median of both inward- and outward-moving orbits, which overlap because of the collisionless nature of dark matter. 
Consequently, within $\mathrm{R_{inf}}$, most of the mass has already passed through the cluster, having completed at least one orbit. Beyond $\mathrm{R_{inf}}$, the majority of the mass is undergoing its first infall into the cluster and $\mathrm{v_r}$ increases until it reaches zero again at $\mathrm{R_{ta}}$. This turnaround suggests that galaxies at larger radii are slowing down, marking the boundary of the infall region. 

Beyond the infall region ($\mathrm{r} \gtrsim 4.5~\mathrm{R_{200}}$), galaxies exhibit positive radial velocities, consistent with the Hubble expansion. At this large distance the behavior of $\langle \mathrm{v_r} \rangle$ indicates galaxies transitioning to the large-scale environment, where their motion is primarily influenced by cosmic expansion rather than gravitational binding to the cluster. In this regime, the Hubble flow from the expanding universe becomes the dominant.

In the middle and right panels, the tangential velocity components, $\mathrm{v_\theta}$ and $\mathrm{v_\phi}$, exhibit an almost uniform distribution throughout the cluster, with median values close to zero. In contrast to the radial component, the tangential velocities show no marked infall or expansion features, as indicated by the smooth and consistent contours in their distribution. Note that any significant deviations from zero would indicate non-relaxed dynamics or asymmetric galaxy distributions.

\subsection{Velocity Anisotropy Profile of the entire Galaxy Cluster Sample at $z=0$} \label{sec:full}

In this section, we examine the velocity anisotropy profile, $\beta(r)$, for the entire sample of galaxy clusters in the simulation at redshift $z=0$.  The profile is explored from the cluster center ($\mathrm{r} = 0$) out to ($\mathrm{r} = 10~\mathrm{R_{200}}$), encompassing the kinematic behavior of galaxies across the virialized core, the infall region, and the outer regions influenced by the Hubble flow. The sample includes a total of $\mathrm{N_c} = 297,357$ clusters, providing a statistically significant dataset. This large sample size enhances the robustness and reliability of our analysis, enabling more accurate insights into the dynamical behavior of galaxy clusters.

Figure~\ref{fig:betaall} presents the results in two panels. The left panel shows the radial velocity dispersion (\(\sigma_r\)), polar velocity dispersion (\(\sigma_\theta\)), and azimuthal velocity dispersion (\(\sigma_\phi\)) as functions of the normalized radius, \(\mathrm{r}/\mathrm{R_{200}}\). These velocity dispersions are normalized by the characteristic virial velocity \(\mathrm{V}_{200}\). The right panel presents the velocity anisotropy profile, \(\beta(r)\), as defined in equation \ref{eq:beta}. Note that the tangential velocity dispersions, \(\sigma_\theta\) and \(\sigma_\phi\), remain nearly equal throughout the radial range, indicating no significant preferred tangential direction \citep{Carlberg97,Lotz19}.

Within \(\mathrm{r} \lesssim 1.7~\mathrm{R_{200}}\), all three velocity dispersion components remain relatively high, reflecting the random, nearly isotropic motions characteristic of the virialized core of the cluster \citep{Binney87,Biviano13}. In this region, the radial velocity dispersion, \(\sigma_r\) (red line), is slightly higher than the tangential components, \(\sigma_\theta\) (blue line) and \(\sigma_\phi\) (purple dashed line), indicating a mild preference for radial motions even near the cluster center \citep{Merritt85,Wojtak09}. 
Moving outward, \(\sigma_r\) decreases steadily, while both tangential components decline more steeply, leading to an increase in the velocity anisotropy parameter, \(\beta(r)\). Near the cluster center, \(\beta(r)\) starts at a relatively low value of about 0.2, indicating a near-isotropic velocity distribution. As \(\mathrm{r}/\mathrm{R_{200}}\) increases, \(\beta(r)\) rises steadily, reaching a peak value of approximately 0.5 at \(\mathrm{r} \simeq 1.7~\mathrm{R_{200}}\), marking a transition to more radially dominated orbits in the region. 
This peak is consistent with previous simulation studies, such as those by \citet{Cuesta08,Iannuzzi12}, which attribute it to the influence of infalling galaxies.

\begin{figure*}
    \centering \includegraphics[width=1\linewidth]{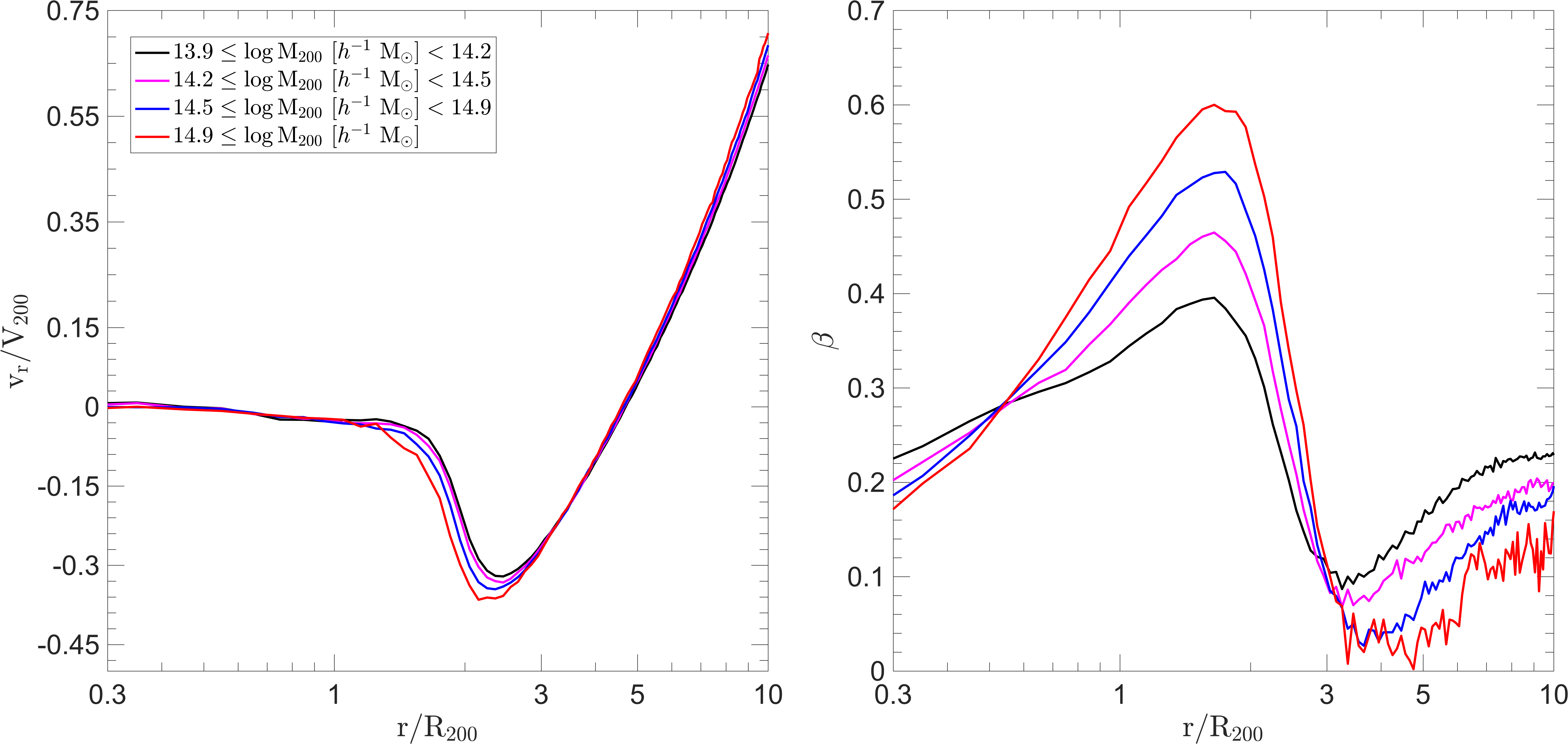}
    \caption{
    Velocity anisotropy profile $\beta$ as a function of cluster mass at redshift $z=0$. The left panel shows the normalized radial velocity, \( \mathrm{v_r} / V_{200} \), as a function of the normalized radius, \( \mathrm{r/ R_{200}}\), for four cluster mass bins: \( 13.9 \leq \log{\mathrm{M_{200}}} [h^{-1} M_\odot] < 14.2 \) (black), \( 14.2 \leq \log{\mathrm{M_{200}}} [h^{-1} M_\odot] < 14.5 \) (magenta), \( 14.5 \leq \log{\mathrm{M_{200}}} [h^{-1} M_\odot] < 14.9 \) (blue), and \( 14.9 \leq \log{\mathrm{M_{200}}} [h^{-1} M_\odot] \) (red). The right panel shows the corresponding velocity anisotropy profiles, \(\beta(r)\), as a function of \(\mathrm{r / R_{200}} \), for the same mass bins. The profiles reveal the dependence of both \( \mathrm{v_r} / V_{200} \) and \(\beta(r)\) on cluster mass and radius, highlighting differences in the dynamical structure of low- and high-mass clusters.
    }
    \label{fig:beta_mass}
\end{figure*}

In the region \(1.7~\mathrm{R_{200}} \lesssim \mathrm{r} \lesssim 3.4~\mathrm{R_{200}}\), while \(\sigma_r\) continues to gradually decrease, reaching its minimum value at \(\approx 3.4~\mathrm{R_{200}}\), both \(\sigma_\theta\) and \(\sigma_\phi\) begin to rise, reaching a maximum around \(3.4~\mathrm{R_{200}}\). This increase in tangential velocity dispersions is attributed to the overlap of inward- and outward-moving orbits, where some galaxies have completed at least one pericentric passage and are now moving outward, while others continue infalling. The coexistence of these orbital populations leads to angular momentum redistribution, enhancing tangential velocity components. As a result, \(\beta(r)\) decreases gradually, reaching lower values at \(\mathrm{r} \approx 3.4~\mathrm{R_{200}}\). This decline is driven by the rise in tangential velocity dispersions (\(\sigma_\theta\) and \(\sigma_\phi\)), as observed in the left panel, indicating a transition to more mixed orbits where tangential motions become more significant.

Beyond \(\mathrm{r} \gtrsim 3.4~\mathrm{R_{200}}\), \(\sigma_r\) begins to increase again, while \(\sigma_\theta\) and \(\sigma_\phi\) flatten and remain relatively constant. This marks the transition to a region dominated by first-infalling galaxies that have not yet completed an orbit within the cluster. Their motion remains predominantly radial, as they experience minimal angular momentum redistribution due to limited dynamical interactions with the cluster environment. At these radii, the contribution of first-infalling galaxies becomes more significant, and the transition to the cosmic environment is increasingly governed by large-scale structure dynamics and the Hubble flow \citep{Biviano04, Abdullah13, Lemze12}. As a result, \(\beta(r)\) starts to increase beyond \(\mathrm{r} \gtrsim 3.4~\mathrm{R_{200}}\), reflecting the growing dominance of radial orbits in the outskirts of the cluster.

\subsection{Velocity Anisotropy Profile as a Function of Cluster Mass}
In this section, we examine the velocity anisotropy profile, \(\beta(r)\), as a function of cluster mass at redshift \(z=0\). By dividing the cluster sample into four mass bins, we analyze how \(\beta(r)\) varies with radius to study the kinematic behavior of galaxies both within and beyond the virial radius. This approach allows us to assess the dynamical state of galaxy clusters across different mass ranges, providing insights into the mass-dependent evolution of their orbital structure.

The left panel of Figure \ref{fig:beta_mass} shows the normalized radial velocity profiles, \(\mathrm{v_r}/\mathrm{V}_{200}\), as a function of \(\mathrm{r/R_{200}}\) for the four cluster mass bins: \( 13.9 \leq \log \mathrm{M}_{200}\) [\hm] \(< 14.2 \) (black), \( 14.2 \leq \log \mathrm{M}_{200}~[h^{-1} ~\mathrm{M}_\odot] < 14.5 \) (magenta), \( 14.5 \leq \log \mathrm{M}_{200}~[h^{-1}~ \mathrm{M}_\odot] < 14.9 \) (blue), and \( 14.9 \leq \log \mathrm{M}_{200}~[h^{-1} ~\mathrm{M}_\odot] \) (red). 
The plot shows that the hydrostatic radius, \(\mathrm{R_{hs}}\), is smaller in massive clusters (\(\mathrm{R_{hs}} \approx 1.3~ \mathrm{R}_{200}\)) compared to low-mass clusters (\(\mathrm{R_{hs}} \approx 1.5 ~\mathrm{R}_{200}\)). This difference arises because low-mass clusters, with shallower gravitational potential wells, have shorter dynamical timescales, given by \(\mathrm{t_{{dyn}}} = \sqrt{\mathrm{r^3 / G M(r)}}\). These shorter timescales allow low-mass clusters to reach dynamical equilibrium more quickly, enabling the virialized region to extend farther relative to their total size.
In contrast, massive clusters require more time to relax and virialize due to their longer dynamical timescales. The virialized region therefore forms a smaller fraction of the total cluster size. 

\begin{figure*}
    \centering
    \includegraphics[width=1\linewidth]{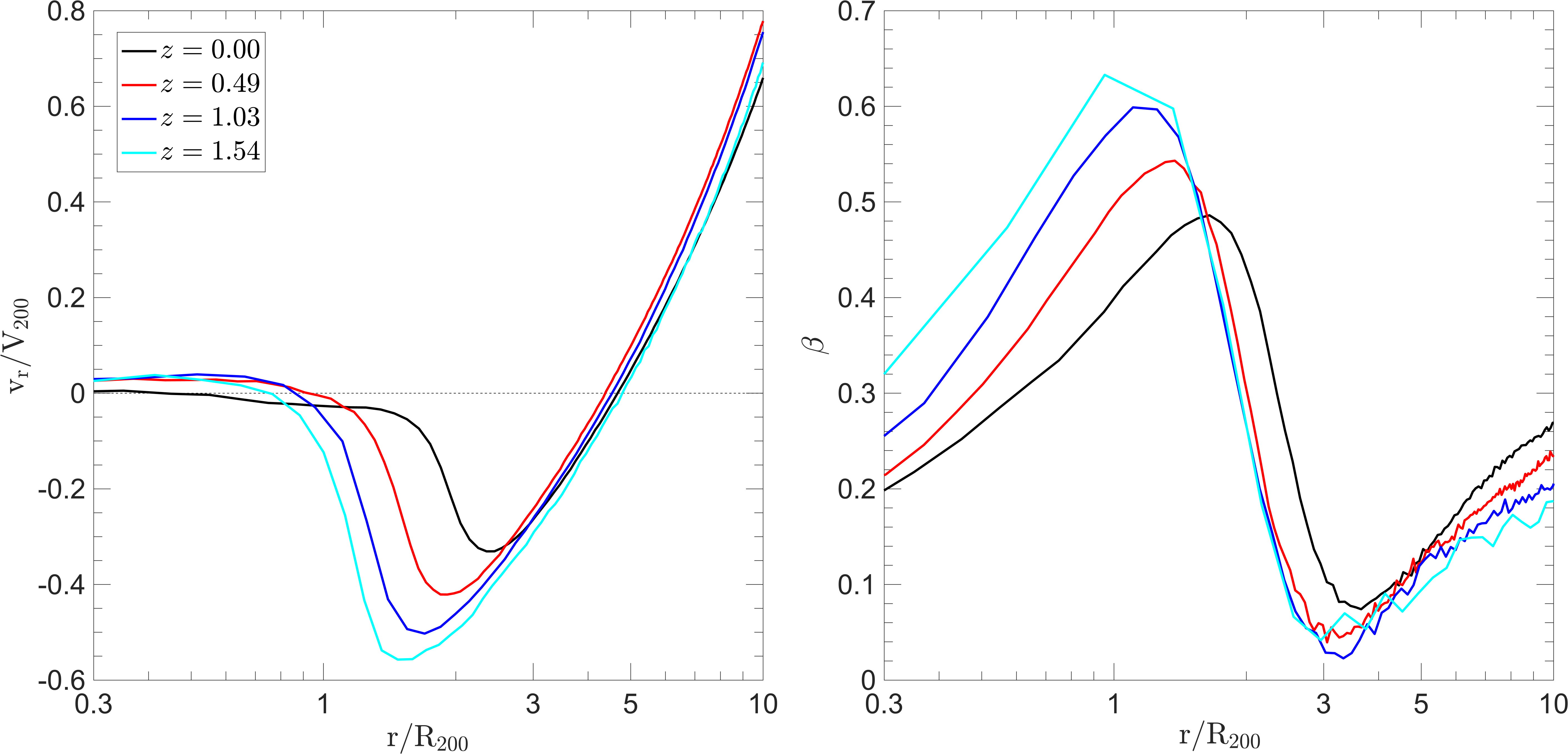}
    \caption{
        Velocity anisotropy profile $\beta$ as a function of redshift. The left panel shows the normalized radial velocity, \( \mathrm{v_r} / \mathrm{V_{200}} \), as a function of the normalized radius, \(\mathrm{ r / R_{200}} \), at redshifts $z = [0.00, 0.49, 1.03, 1.54]$ as shown in the legend. The right panel shows the corresponding velocity anisotropy profiles, \(\beta(r)\), as a function of \(\mathrm{ r / R_{200}} \), for the same redshifts. The profiles reveal the dependence of both \( \mathrm{v_r / V_{200}} \) and \(\beta(r)\) on redshift and radius, highlighting differences in the dynamical structure of clusters at different redshifts.
    }
    \label{fig:beta_redshift}
\end{figure*}

Furthermore, the plot shows that the infall region of low-mass clusters is smaller than that of high-mass clusters, reflecting the shallower gravitational potential wells of smaller systems. The absolute value of the minimum radial velocity, \(|\mathrm{v_r}|\), is lower in low-mass clusters compared to high-mass clusters, indicating a weaker gravitational pull. Interestingly, the turnaround radius, \(\mathrm{R_{ta}}\), which marks the outer boundary of the infall region, scales with the virial radius as \(\mathrm{R_{ta}} \approx 4.5 ~\mathrm{R_{200}}\) across different cluster mass ranges.

\begin{figure*}
    \centering
    \includegraphics[width=1\linewidth]{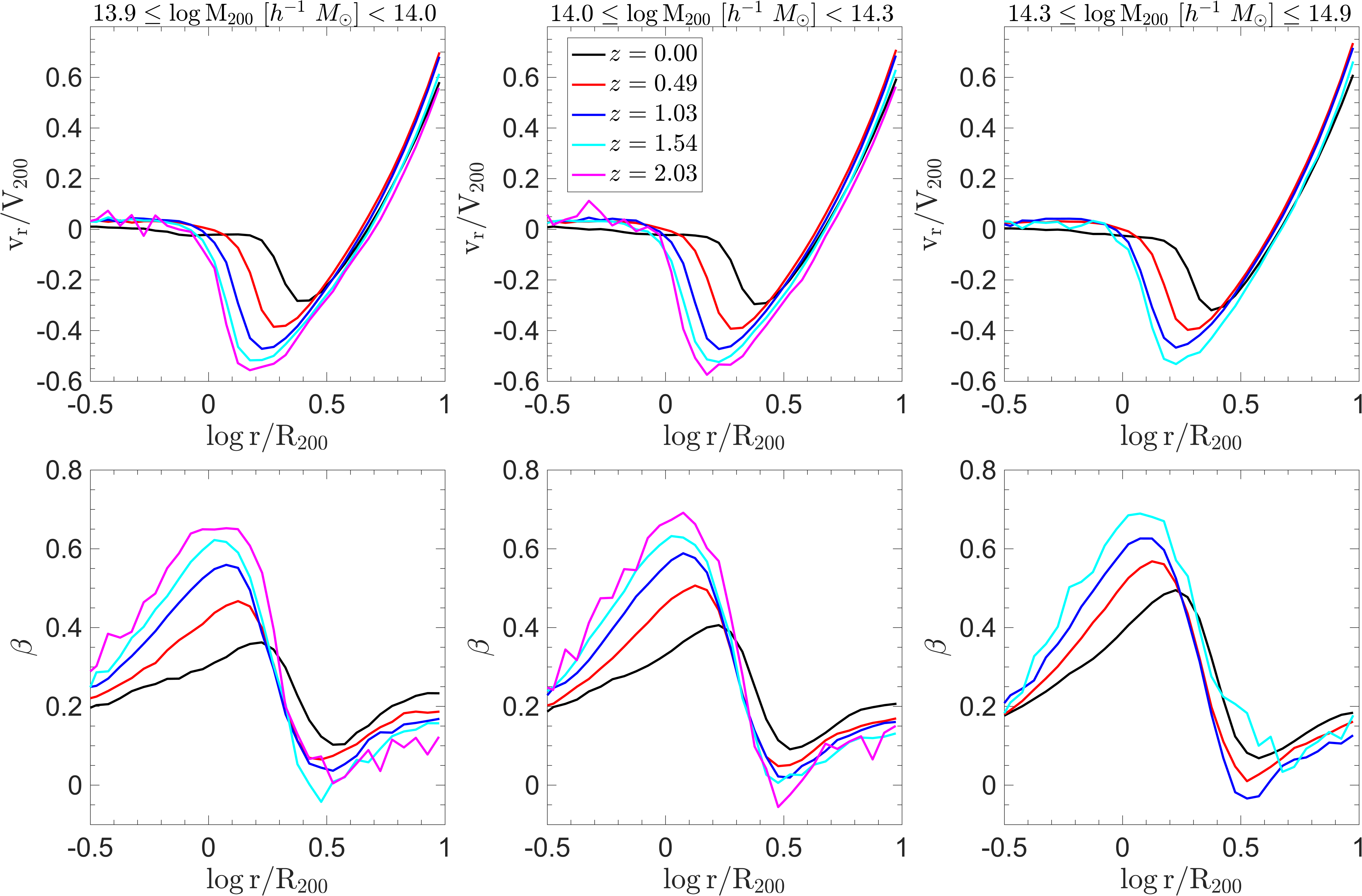}
    \caption{Radial velocity and velocity anisotropy profiles of galaxies in galaxy clusters as a function of redshift and cluster mass. 
        The top row shows the mean radial velocity normalized to \( \mathrm{V_{200}} \) as a function of normalized cluster-centric radius (\(\mathrm{ r/R_{200}} \)) for different redshifts (\( z = 0.00, 0.49, 1.03, 1.54 \)) and cluster mass bins. 
        The bottom row displays the corresponding velocity anisotropy parameter, \( \beta(r) \), which quantifies the orbital anisotropy of galaxies.
    }
    \label{fig:beta_mass_redshift}
\end{figure*}

The right panel shows the velocity anisotropy profiles, \(\beta(r)\), as a function of \( \mathrm{r / R_{200}} \). In the inner regions, \(\beta(r)\) increases steadily, starting from values around 0.2, indicating the presence of nearly isotropic orbits. This rise reflects the growing dominance of radial orbits as galaxies move outward into the infall region (as we discussed in Section \ref{sec:full}). The peak in \(\beta(r)\) occurs at approximately \( \mathrm{r} \approx 1.7~ \mathrm{R_{200}} \), with the peak value increasing with cluster mass, i.e., more massive clusters exhibit a steeper increase in $\beta$.
Low-mass clusters exhibit a shallow peak in radial anisotropy ($\beta_{\mathrm{max}} \approx 0.4$ at $\mathrm{r} \approx 1.7\mathrm{R_{200}}$), while most massive clusters show a pronounced maximum ($\beta_{\mathrm{max}} \approx 0.6$ at $r\approx 1.7\mathrm{R_{200}}$). This suggests that larger clusters exhibit greater radial anisotropy as a result of their stronger gravitational influence, characterized by deeper potential wells. These deeper potentials enhance radial infall, as they more effectively pull matter along radial trajectories, leading to a more pronounced dominance of radial motions over tangential ones. 

Beyond the peak, \(\beta(r)\) decreases toward \( \mathrm{r} \approx 3\mathrm{R_{200}} \) due to the overlapping of inward- and outward-moving orbits as galaxies complete their first apocenter passage. This orbital mixing reduces the dominance of radial orbits and enhances tangential motions. For \( \mathrm{r} > 3\mathrm{R_{200}} \), \(\beta(r)\) begins to rise again, reflecting the influence of first-infalling galaxies with predominantly radial orbits (see Section \ref{sec:full}). 

These results highlight the dependence of galaxy kinematics and orbital anisotropy on cluster mass and radial position. High-mass clusters exhibit more pronounced anisotropy peaks and larger outer infall regions, emphasizing the role of mass in defining both the extent of the stable region and the dynamics of the infall zone. In contrast, low-mass clusters maintain more localized dynamics. Overall, these trends are consistent with hierarchical structure formation, where massive clusters dominate their surroundings through extended gravitational influence, offering key insights into the dynamical assembly of clusters across different mass regimes.
\setlength{\tabcolsep}{5pt}
\subsection{Velocity Anisotropy Profile as a Function of redshift}
In this section, we examine the evolution of the velocity anisotropy profile, \(\beta(r)\), with redshift. The analysis includes five snapshots at redshifts \(z = [0.00, 0.49, 1.03, 1.54]\). Understanding the evolution of \(\beta(r)\) provides insights into the dynamical state of galaxies within these clusters. By studying the trends in these profiles across different redshifts, we aim to understand how the kinematic properties of galaxies in clusters change over cosmic time and how this evolution depends on cluster mass.

The left panel of Figure~\ref{fig:beta_redshift} shows the evolution of the normalized radial velocity profiles, $\mathrm{v_r}/\mathrm{V}_{200}$, as a function of $\mathrm{r/R_{200}}$ with redshift. The hydrostatic radius increases from $\mathrm{R_{hs}}\approx 0.9\mathrm{R_{200}}$ at $z=1.54$ to $\mathrm{R_{hs}}\approx 1.4\mathrm{R_{200}}$ at $z=0$, indicating longer dynamical timescales at lower redshifts. This trend arises because galaxy clusters gradually become more relaxed, virialized, and attain hydrostatic equilibrium as redshift decreases. 
At higher redshifts, clusters exhibit more negative radial velocities, indicative of higher mass accretion rates \citep{Wetzel15}. The absolute value of the minimum radial velocity decreases from $\sim0.33$ at $z=0$ to $\sim0.56$ at $z=1.54$. The infall radius $\mathrm{R_{inf}}$ increases with decreasing redshift. The infall radius increases from $\mathrm{R_{inf}}\approx 1.5\mathrm{R_{200}}$ at $z=1.54$ to $\mathrm{R_{inf}}\approx 2.4\mathrm{R_{200}}$ at $z=0$. This is consistent with the results obtained by \cite{Wetzel15}. These trends arise due to the decline in both the cosmic accretion rate and the density growth of the cluster over cosmic time \citep{Wetzel15}.
The scaled turnaround radius ($\mathrm{R_{ta}}/\mathrm{R_{200}}$) changes slightly across all redshifts, where $\mathrm{R_{ta}}\approx4.8\mathrm{R_{200}}$ at $z=1.54$ versus $\mathrm{R_{ta}}\approx4.7\mathrm{R_{200}}$ at $z=0$, suggesting that the dependence of $\mathrm{R_{ta}}/\mathrm{R_{200}}$ on the redshift is negligible for galaxy clusters (similar to results obtained by \cite{Lau15}).

\begin{table*}
\centering
\caption{Best-fit parameters \( a \) and \( b \) for the scaling relation \( \log \mathrm{R} = a \log(\mathrm{M}_{200}/\mathrm{M_{piv}}) + b \) at different redshifts for the hydrostatic radius \( \mathrm{R}_{\mathrm{hs}} \), infall radius \( \mathrm{R}_{\mathrm{inf}} \), and turnaround radius \( \mathrm{R}_{\mathrm{ta}} \), where \(\mathrm{M_{piv}}=5\times10^{14}~h^{-1}~ \mathrm{M}_{\odot}\) (Section \ref{sec:scalingrelation}). The reduced chi-squared values \( \chi^2_\mathrm{red} \) are also listed for each fit.}
\begin{tabular}{cc}
\hline
Fit &  $\chi^2_\mathrm{{red}}$ \\
\hline
\multicolumn{2}{c} {$z=0.00$}\\
\hline
$\log{\mathrm{R_{hs}}} =(0.278\pm0.008) \log{(\mathrm{M}_{200}/\mathrm{M_{piv}})} + (0.228\pm0.003)$&$6.57\times 10^{-4}$\\

$\log{\mathrm{R_{inf}}}=(0.290\pm0.001) \log{(\mathrm{M}_{200}/\mathrm{M_{piv}})} + (0.488\pm0.001)$&$6.49\times 10^{-6}$\\

$\log{\mathrm{R_{ta}}} =(0.324\pm0.001) \log{(\mathrm{M}_{200}/\mathrm{M_{piv}})} + (0.777\pm0.001)$&$1.46\times 10^{-6}$\\
\hline

\multicolumn{2}{c} {$z=0.49$}\\
\hline
$\log{\mathrm{R_{hs}}} =(0.330\pm0.010) \log{(\mathrm{M}_{200}/\mathrm{M_{piv}})} + (0.079\pm0.004)$&$1.60\times 10^{-3}$\\

$\log{\mathrm{R_{inf}}}=(0.286\pm0.003) \log{(\mathrm{M}_{200}/\mathrm{M_{piv}})} + (0.310\pm0.001)$&$4.55\times 10^{-5}$\\

$\log{\mathrm{R_{ta}}} =(0.319\pm0.001) \log{(\mathrm{M}_{200}/\mathrm{M_{piv}})} + (0.667\pm0.001)$&$3.44\times 10^{-6}$\\
\hline

\multicolumn{2}{c} {$z=1.03$}\\
\hline
$\log{\mathrm{R_{hs}}} =(0.378\pm0.013) \log{(\mathrm{M}_{200}/\mathrm{M_{piv}})} - (0.069\pm0.006)$&$1.50\times 10^{-3}$\\

$\log{\mathrm{R_{inf}}}=(0.295\pm0.011) \log{(\mathrm{M}_{200}/\mathrm{M_{piv}})} + (0.149\pm0.003)$&$1.30\times 10^{-3}$\\

$\log{\mathrm{R_{ta}}} =(0.313\pm0.006) \log{(\mathrm{M}_{200}/\mathrm{M_{piv}})} + (0.583\pm0.003)$&$5.39\times 10^{-5}$\\
\hline

\multicolumn{2}{c} {$z=1.54$}\\
\hline
$\log{\mathrm{R_{hs}}} =(0.432\pm0.059) \log{(\mathrm{M}_{200}/\mathrm{M_{piv}})} - (0.163\pm0.031)$&$2.10\times 10^{-3}$\\

$\log{\mathrm{R_{inf}}}=(0.335\pm0.009) \log{(\mathrm{M}_{200}/\mathrm{M_{piv}})} + (0.073\pm0.009)$&$6.50\times 10^{-4}$\\

$\log{\mathrm{R_{ta}}} =(0.329\pm0.008) \log{(\mathrm{M}_{200}/\mathrm{M_{piv}})} + (0.533\pm0.005)$&$4.05\times 10^{-5}$\\
\hline
\end{tabular}
\label{tab:radius_values}
\end{table*}

The right panel of Figure~\ref{fig:beta_redshift} illustrates the evolution of the velocity anisotropy profile \(\beta(r)\) with redshift, using the same redshift snapshots as in the left panel. \(\beta(r)\) profiles exhibit similar general behavior across all redshifts. \(\beta(r)\) increases until it reaches its peak (see Section \ref{sec:full}). The peak of \(\beta(r)\) decreases as the redshift decreases. Specifically, the peak value decreases from approximately ($\beta_{\mathrm{max}} \approx 0.63$ at $\mathrm{r} \approx 1\mathrm{R_{200}}$ and $z=1.54$) to ($\beta_{\mathrm{max}} \approx 0.5$ at $\mathrm{r} \approx 1.7\mathrm{R_{200}}$ and $z=0$). This trend suggests a decrease in radial anisotropy at lower redshifts, reflecting the transition of galaxy clusters from a highly radially anisotropic state at high redshift to a more isotropic velocity distribution over cosmic time. 
Beyond its maximum, $\beta$ decreases with increasing distance from the cluster center, reaching a minimum. This minimum value increases with decreasing redshift, changing from ($\beta_{\mathrm{min}} \approx 0.04$ at $\mathrm{r} \approx 3\mathrm{R_{200}}$ and $z = 1.54$) to ($\beta_{\mathrm{min}} \approx 0.08$ at $\mathrm{r} \approx 3.6\mathrm{R_{200}}$ and $z = 0$).
Beyond this minimum, $\beta$ increases again, asymptotically approaching the value associated with the Hubble flow, reflecting the influence of the expanding universe at large radii.

 \begin{figure*} 
    \centering  \includegraphics[width=1\linewidth]{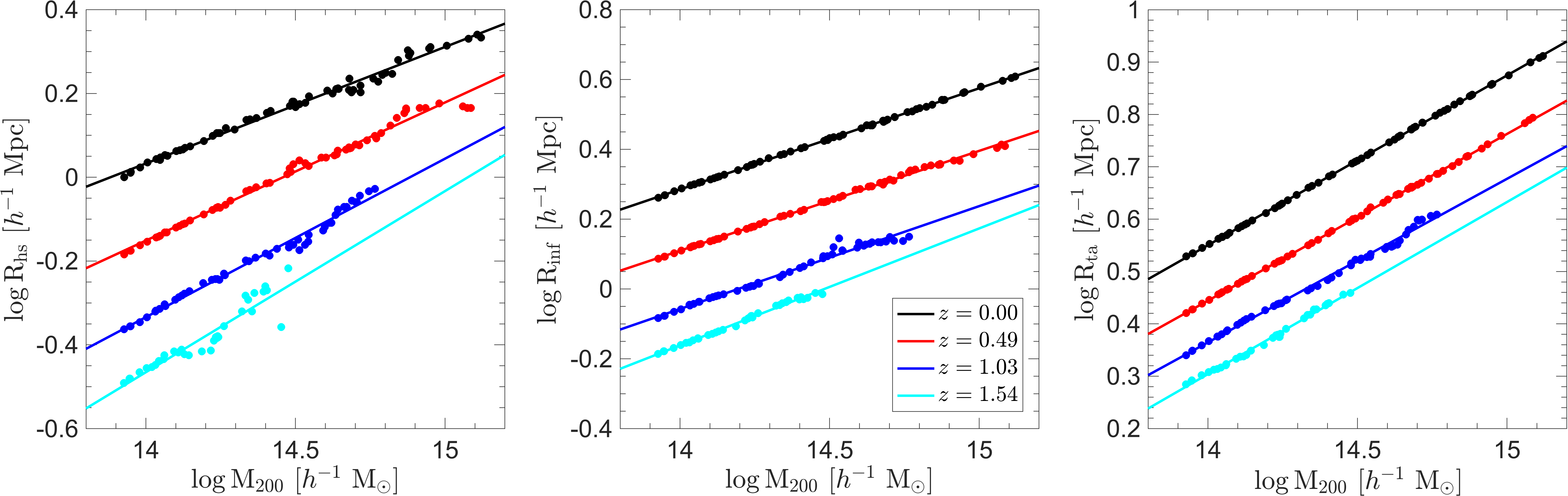}
\caption{Scaling relations between cluster mass \( \log{\mathrm{M_{200}}} \) and the hydrostatic radius \( \log \mathrm{R}_{\mathrm{hs}} \) (left), infall radius \( \log \mathrm{R}_{\mathrm{inf}} \) (middle), and turnaround radius \( \log \mathrm{R}_{\mathrm{ta}} \) (right) at four different redshifts: \( z = 0.00 \) (black), \( z = 0.49 \) (red), \( z = 1.03 \) (blue), and \( z = 1.54 \) (cyan). Solid lines represent the best-fit power-law relations at each redshift.}
\label{fig:scalingrelation}
\end{figure*}

To investigate the redshift evolution of galaxy clusters as a function of mass, we divide the cluster sample into three mass bins: 
\( 13.9 \leq \log \mathrm{M}_{200}~[h^{-1}~ \mathrm{M}_{\odot}] < 14.0 \), 
\( 14.0 \leq \log \mathrm{M}_{200}~[h^{-1}~ \mathrm{M}_{\odot}] < 14.3 \), and 
\( 14.3 \leq \log \mathrm{M}_{200}~[h^{-1}~ \mathrm{M}_{\odot}] \leq 14.9 \). 
For each mass bin, we analyze the redshift evolution of the radial velocity profile and the velocity anisotropy parameter \( \beta(r) \), as presented in Figure~\ref{fig:beta_mass_redshift}. The results indicate that the dynamical state of galaxy clusters with similar masses varies significantly with redshift, highlighting the ongoing process of structure formation. Clusters at high redshift are dynamically younger, still undergoing active mass assembly, while clusters at low redshift are more relaxed and closer to virial equilibrium.

\begin{figure}
    \centering  \includegraphics[width=0.9\linewidth]{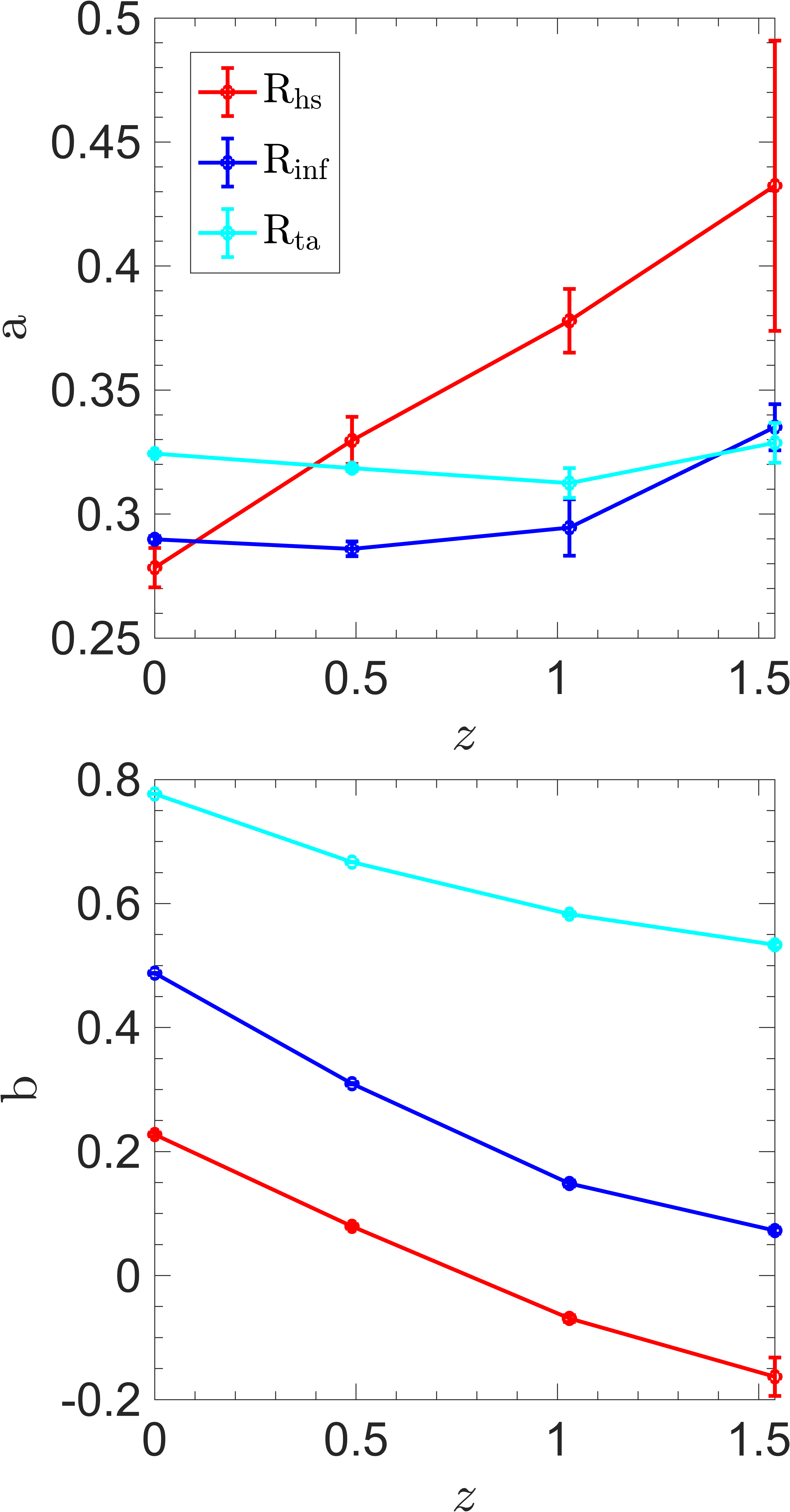}
    \caption{Redshift evolution of the best-fit parameters \( a \) (top panel) and \( b \) (bottom panel) in the scaling relation  \( \log \mathrm{R} = a \log(\mathrm{M_{200}/M_{\mathrm{piv}})} + b \), with \(\mathrm{M}_{\mathrm{piv}} = 5 \times 10^{14}~h^{-1}~\mathrm{M_\odot} \). The three curves correspond to the hydrostatic radius \( \mathrm{R_{hs}} \) (red), infall radius \( \mathrm{R_{inf}} \) (blue), and turnaround radius \( \mathrm{R_{ta}} \) (cyan). While the slope \( a \) increases with redshift for \( \mathrm{R_{hs}} \), it remains nearly constant for \( \mathrm{R_{inf}} \) and \( \mathrm{R}_{\mathrm{ta}} \). The normalization \( b \) shows a steady decline with redshift for all three radii.
}
\label{fig:slopenorm_evolution}
\end{figure}

At high redshift, clusters exhibit stronger infall motions, suggesting that they are still in the process of accreting galaxies from their surroundings. The velocity anisotropy profile at these redshifts shows that galaxies predominantly follow radial orbits, indicating that many galaxies are infalling for the first time and have not yet undergone significant interactions or mixing within the cluster potential. Additionally, the transition to isotropy occurs at larger radii, implying that the outskirts of high-redshift clusters are dynamically less mature, with ongoing accretion shaping the kinematic structure.

In contrast, clusters at low redshift appear more dynamically evolved. The reduced infall motions indicate that a significant fraction of galaxies has already virialized within the cluster potential. The velocity anisotropy parameter shows a smoother profile, with a weaker radial bias, leading to a more isotropic velocity distribution over time. Furthermore, the hydrostatic radius increases with decreasing redshift. This is because galaxy clusters gradually become more relaxed and virialized as redshift decreases.

\subsection{Evolution of the Scaling Relations Between \(\mathrm{M_{200}}\) and Key Cluster Radii}
\label{sec:scalingrelation}

In this section, we explore the relationship between the hydrostatic ($\mathrm{R_{hs}}$), infall ($\mathrm{R_{inf}}$), and turnaround ($\mathrm{R_{ta}}$) radii and the cluster mass (\( \mathrm{M_{200}} \)). These radii define key physical boundaries within galaxy clusters (see Section~\ref{sec:full} for definitions of these radii).
To quantify these relationships, we first determine the values of these radii for clusters in different mass bins at various redshifts. Both \( \mathrm{R_{hs}} \) and \( \mathrm{R_{ta}} \) correspond to regions where \( \mathrm{v/V_{200}} \approx 0 \), while \( \mathrm{R_{inf}} \) corresponds to the location of minimum radial velocity. We follow the method of \citet{Busha05} to determine \( \mathrm{R_{hs}} \), as it is challenging to identify directly. Starting from the cluster center, we define \( \mathrm{R_{hs}} \) as the smallest radius at which the mean radial velocity satisfies $\langle \mathrm{v_r\rangle/ V_{200}} = -0.04$.

We then derive the scaling relations between each radius and mass \( \mathrm{M_{200}} \) at each redshift by fitting a power-law relation of the form:
\begin{equation}
    \log \mathrm{R} = a \log{\mathrm{M_{200}}} + b,
\end{equation}

\noindent where \( \mathrm{R} \) represents \( \mathrm{R}_{\mathrm{hs}} \), \( \mathrm{R}_{\mathrm{inf}} \), or \( \mathrm{R}_{\mathrm{ta}} \), and \( a \) and \( b \) are the best-fit parameters that vary with redshift.

By fitting this relation at different redshifts, we trace the evolution of these characteristic scales over cosmic time. The best-fit values of \( a \) and \( b \) for each redshift are summarized in Table~\ref{tab:radius_values}, providing insights into how these radii scale with mass and how their dependence on \( \mathrm{M_{200}} \) evolves.

Figure~\ref{fig:scalingrelation} shows the scaling relations between \( \log{\mathrm{M_{200}}} \) and the three characteristic radii: \( \log \mathrm{R_{hs}} \) (left panel), \( \log \mathrm{R_{inf}} \) (middle panel), and \( \log \mathrm{R_{ta}} \) (right panel), at four different redshifts: $ z = [0.00, 0.49, 1.03, 1.54]$. In each panel, clusters are binned by mass, and the corresponding radii are plotted with best-fit power-law relations overlaid as solid lines. The figure illustrates that all three radii increase with mass and decrease with redshift. Additionally, the slopes and normalizations of the fits vary systematically with redshift, reflecting the evolving dynamical structure of galaxy clusters over cosmic time.

Figure~\ref{fig:slopenorm_evolution} illustrates the redshift evolution of the best-fit parameters in the scaling relation \( \log\mathrm{R} = a \log(\mathrm{M_{200}/M_{\mathrm{piv}})} + b \), where \( \mathrm{M}_{\mathrm{piv}} = 5 \times 10^{14}~h^{-1}~\mathrm{M_\odot} \). The upper panel shows the evolution of the slope \( a \), while the lower panel shows the evolution of the normalization \( b \). We find that the slope \( a \) increases significantly with redshift for \( \mathrm{R_{hs}} \), while remaining nearly constant for both \( \mathrm{R_{inf}} \) and \( \mathrm{R_{ta}} \). In contrast, the normalization \( b \) decreases with increasing redshift for all three radii, reflecting the overall shrinking of cluster boundaries at fixed mass as we move to earlier cosmic times.
\section{Conclusion} \label{sec:conc}
In this study, we investigated the velocity anisotropy profile, $\beta(r)$, of galaxy clusters using the Uchuu-UniverseMachine mock galaxy catalog ~\citep{Ishiyama21, Aung23}. The large volume ($8\,h^{-3}\,\mathrm{Gpc}^3$) and high resolution  ($3.27 \times 10^8 \, h^{-1} \, \mathrm{M_\odot}$ particle mass) of the Uchuu simulation enabled us to explore galaxy cluster dynamics with both statistical power and spatial detail across a wide range of masses and redshifts. Leveraging this dataset, we examined how $\beta(r)$ varies with cluster mass and redshift, and we derived redshift-dependent scaling relations between cluster mass and key physical radii. Our main findings are summarized as follows:

\begin{itemize}[left=0pt,labelsep=0.5em]
    \item Universal shape of $\beta(r)$: The anisotropy profile exhibits a consistent radial trend across all clusters—starting from nearly isotropic values ($\beta \approx 0.2$) in the core, peaking at $\sim1.7~\mathrm{R_{200}}$ ($\beta \approx 0.5$–$0.6$), declining near $\sim3.4~\mathrm{R_{200}}$ due to orbital mixing, and rising again in the outskirts where first-infalling galaxies dominate.

    \item Dependence on cluster mass: More massive clusters show higher peak $\beta$ values, indicating stronger radial anisotropy and deeper potential wells. The transition from virialized to infall regions is also sharper in high-mass clusters.

    \item Redshift evolution: Clusters at higher redshifts exhibit stronger infall velocities and more radially dominated orbits. With cosmic time, clusters become dynamically relaxed, resulting in a more isotropic orbital distribution and an outward shift in the hydrostatic and infall radii.

    \item Scaling relations: We derived redshift-dependent power-law relations between cluster mass ($\mathrm{M_{200}}$) and three key radii: hydrostatic ($\mathrm{R}_{\mathrm{hs}}$), infall ($\mathrm{R}_{\mathrm{inf}}$), and turnaround ($\mathrm{R}_{\mathrm{ta}}$). These relations evolve with redshift and reflect the changing dynamical structure of clusters over time.


    \item Implications: Our results provide a robust theoretical framework for interpreting cluster dynamics in both simulations and observations. This work supports future spectroscopic surveys such as Subaru PFS, DESI and Euclid and contributes to the broader understanding of structure formation and cosmology.
\end{itemize}

Future work can extend this analysis in several directions. First, it would be valuable to compare our simulation-based results with observational measurements of $\beta(r)$ from spectroscopic surveys such as PFS, DESI and Euclid, enabling a direct test of the theoretical predictions presented here. Incorporating galaxy properties such as morphology, color, or stellar mass could also reveal how orbital anisotropy depends on galaxy type. Furthermore, applying the same analysis to hydrodynamical simulations would allow us to quantify the role of baryonic physics in shaping $\beta(r)$, especially in the core regions where baryonic effects are expected to be most significant. Finally, integrating the anisotropy profile into dynamical mass estimation frameworks may improve cluster mass calibration in both current and future cosmological analyses.

\section*{Acknowledgments}
MA thanks the Instituto de Astrof\'{\i}sica de Andaluc\'{\i}a (IAA-CSIC), Centro de Supercomputaci\'on de Galicia (CESGA), and the Spanish academic and research network (RedIRIS) in Spain for hosting Uchuu DR1 in the Skies \& Universes site (\url{http://www.skiesanduniverses.org/}) for cosmological simulations.  The Uchuu simulations were carried out on the Aterui II supercomputer at the Center for Computational Astrophysics, CfCA, of the National Astronomical Observatory of Japan, and the K computer at the RIKEN Advanced Institute for Computational Science. GW gratefully acknowledges support from the National Science Foundation through grant AST-2205189.
TI has been supported by IAAR Research Support Program in Chiba
University Japan, MEXT/JSPS KAKENHI (Grant Number JP19KK0344 and
JP23H04002), MEXT as ``Program for Promoting Researches on the
Supercomputer Fugaku'' (JPMXP1020230406 and JPMXP1020230407), and
JICFuS.

\bibliography{Ref}{}

\begin{thebibliography}{}
\expandafter\ifx\csname natexlab\endcsname\relax\def\natexlab#1{#1}\fi
\providecommand{\url}[1]{\href{#1}{#1}}
\providecommand{\dodoi}[1]{doi:~\href{http://doi.org/#1}{\nolinkurl{#1}}}
\providecommand{\doeprint}[1]{\href{http://ascl.net/#1}{\nolinkurl{http://ascl.net/#1}}}
\providecommand{\doarXiv}[1]{\href{https://arxiv.org/abs/#1}{\nolinkurl{https://arxiv.org/abs/#1}}}

\bibitem[{{Abdullah} {et~al.}(2011){Abdullah}, {Ali}, {Ismail}, \& {Rassem}}]{Abdullah11}
{Abdullah}, M.~H., {Ali}, G.~B., {Ismail}, H.~A., \& {Rassem}, M.~A. 2011, \mnras, 416, 2027, \dodoi{10.1111/j.1365-2966.2011.19178.x}

\bibitem[{{Abdullah} {et~al.}(2013){Abdullah}, {Praton}, \& {Ali}}]{Abdullah13}
{Abdullah}, M.~H., {Praton}, E.~A., \& {Ali}, G.~B. 2013, \mnras, 434, 1989, \dodoi{10.1093/mnras/stt1145}

\bibitem[{{Abdullah} {et~al.}(2023){Abdullah}, {Wilson}, {Klypin}, \& {Ishiyama}}]{Abdullah23}
{Abdullah}, M.~H., {Wilson}, G., {Klypin}, A., \& {Ishiyama}, T. 2023, \apj, 955, 26, \dodoi{10.3847/1538-4357/ace773}

\bibitem[{{Aguirre Tagliaferro} {et~al.}(2021){Aguirre Tagliaferro}, {Biviano}, {De Lucia}, {Munari}, \& {Garcia Lambas}}]{Tagliaferro21}
{Aguirre Tagliaferro}, T., {Biviano}, A., {De Lucia}, G., {Munari}, E., \& {Garcia Lambas}, D. 2021, \aap, 652, A90, \dodoi{10.1051/0004-6361/202140326}

\bibitem[{{Allen} {et~al.}(2011){Allen}, {Evrard}, \& {Mantz}}]{Allen11}
{Allen}, S.~W., {Evrard}, A.~E., \& {Mantz}, A.~B. 2011, \araa, 49, 409, \dodoi{10.1146/annurev-astro-081710-102514}

\bibitem[{{Aung} {et~al.}(2023){Aung}, {Nagai}, {Klypin}, {Behroozi}, {Abdullah}, {Ishiyama}, {Prada}, {P{\'e}rez}, {L{\'o}pez Cacheiro}, \& {Ruedas}}]{Aung23}
{Aung}, H., {Nagai}, D., {Klypin}, A., {et~al.} 2023, \mnras, 519, 1648, \dodoi{10.1093/mnras/stac3514}

\bibitem[{{Baes} \& {van Hese}(2007)}]{Baes07}
{Baes}, M., \& {van Hese}, E. 2007, \aap, 471, 419, \dodoi{10.1051/0004-6361:20077672}

\bibitem[{{Behroozi} {et~al.}(2019){Behroozi}, {Wechsler}, {Hearin}, \& {Conroy}}]{Behroozi19}
{Behroozi}, P., {Wechsler}, R.~H., {Hearin}, A.~P., \& {Conroy}, C. 2019, \mnras, 488, 3143, \dodoi{10.1093/mnras/stz1182}

\bibitem[{{Behroozi} {et~al.}(2013{\natexlab{a}}){Behroozi}, {Wechsler}, \& {Wu}}]{Behroozi13a}
{Behroozi}, P.~S., {Wechsler}, R.~H., \& {Wu}, H.-Y. 2013{\natexlab{a}}, \apj, 762, 109, \dodoi{10.1088/0004-637X/762/2/109}

\bibitem[{{Behroozi} {et~al.}(2013{\natexlab{b}}){Behroozi}, {Wechsler}, {Wu}, {Busha}, {Klypin}, \& {Primack}}]{Behroozi13b}
{Behroozi}, P.~S., {Wechsler}, R.~H., {Wu}, H.-Y., {et~al.} 2013{\natexlab{b}}, \apj, 763, 18, \dodoi{10.1088/0004-637X/763/1/18}

\bibitem[{{Binney} \& {Tremaine}(1987)}]{Binney87}
{Binney}, J., \& {Tremaine}, S. 1987, {Galactic dynamics}

\bibitem[{{Binney} \& {Tremaine}(2008)}]{Binney08}
---. 2008, {Galactic Dynamics: Second Edition}

\bibitem[{{Biviano} \& {Katgert}(2004)}]{Biviano04}
{Biviano}, A., \& {Katgert}, P. 2004, \aap, 424, 779, \dodoi{10.1051/0004-6361:20041306}

\bibitem[{{Biviano} {et~al.}(2013){Biviano}, {Rosati}, {Balestra}, {Mercurio}, {Girardi}, {Nonino}, {Grillo}, {Scodeggio}, {Lemze}, {Kelson}, {Umetsu}, {Postman}, {Zitrin}, {Czoske}, {Ettori}, {Fritz}, {Lombardi}, {Maier}, {Medezinski}, {Mei}, {Presotto}, {Strazzullo}, {Tozzi}, {Ziegler}, {Annunziatella}, {Bartelmann}, {Benitez}, {Bradley}, {Brescia}, {Broadhurst}, {Coe}, {Demarco}, {Donahue}, {Ford}, {Gobat}, {Graves}, {Koekemoer}, {Kuchner}, {Melchior}, {Meneghetti}, {Merten}, {Moustakas}, {Munari}, {Reg{\H{o}}s}, {Sartoris}, {Seitz}, \& {Zheng}}]{Biviano13}
{Biviano}, A., {Rosati}, P., {Balestra}, I., {et~al.} 2013, \aap, 558, A1, \dodoi{10.1051/0004-6361/201321955}

\bibitem[{{Boselli} {et~al.}(2014){Boselli}, {Voyer}, {Boissier}, {Cucciati}, {Consolandi}, {Cortese}, {Fumagalli}, {Gavazzi}, {Heinis}, {Roehlly}, \& {Toloba}}]{Boselli14}
{Boselli}, A., {Voyer}, E., {Boissier}, S., {et~al.} 2014, \aap, 570, A69, \dodoi{10.1051/0004-6361/201424419}

\bibitem[{{Busha} {et~al.}(2005){Busha}, {Evrard}, {Adams}, \& {Wechsler}}]{Busha05}
{Busha}, M.~T., {Evrard}, A.~E., {Adams}, F.~C., \& {Wechsler}, R.~H. 2005, \mnras, 363, L11, \dodoi{10.1111/j.1745-3933.2005.00072.x}

\bibitem[{{Carlberg} {et~al.}(1997){Carlberg}, {Yee}, \& {Ellingson}}]{Carlberg97}
{Carlberg}, R.~G., {Yee}, H.~K.~C., \& {Ellingson}, E. 1997, \apj, 478, 462, \dodoi{10.1086/303805}

\bibitem[{{Carollo} {et~al.}(1995){Carollo}, {de Zeeuw}, \& {van der Marel}}]{Carollo95}
{Carollo}, C.~M., {de Zeeuw}, P.~T., \& {van der Marel}, R.~P. 1995, \mnras, 276, 1131, \dodoi{10.1093/mnras/276.4.1131}

\bibitem[{{Cuesta} {et~al.}(2008){Cuesta}, {Prada}, {Klypin}, \& {Moles}}]{Cuesta08}
{Cuesta}, A.~J., {Prada}, F., {Klypin}, A., \& {Moles}, M. 2008, \mnras, 389, 385, \dodoi{10.1111/j.1365-2966.2008.13590.x}

\bibitem[{{Dolag} {et~al.}(2015){Dolag}, {Gaensler}, {Beck}, \& {Beck}}]{Dolag15}
{Dolag}, K., {Gaensler}, B.~M., {Beck}, A.~M., \& {Beck}, M.~C. 2015, \mnras, 451, 4277, \dodoi{10.1093/mnras/stv1190}

\bibitem[{{Ebeling} {et~al.}(2014){Ebeling}, {Stephenson}, \& {Edge}}]{Ebeling14}
{Ebeling}, H., {Stephenson}, L.~N., \& {Edge}, A.~C. 2014, \apjl, 781, L40, \dodoi{10.1088/2041-8205/781/2/L40}

\bibitem[{{Gunn} \& {Gott}(1972)}]{Gunn72}
{Gunn}, J.~E., \& {Gott}, III, J.~R. 1972, \apj, 176, 1, \dodoi{10.1086/151605}

\bibitem[{{Hirschmann} {et~al.}(2014){Hirschmann}, {Dolag}, {Saro}, {Bachmann}, {Borgani}, \& {Burkert}}]{Hirschmann14}
{Hirschmann}, M., {Dolag}, K., {Saro}, A., {et~al.} 2014, \mnras, 442, 2304, \dodoi{10.1093/mnras/stu1023}

\bibitem[{{Host} {et~al.}(2009){Host}, {Hansen}, {Piffaretti}, {Morandi}, {Ettori}, {Kay}, \& {Valdarnini}}]{Host09}
{Host}, O., {Hansen}, S.~H., {Piffaretti}, R., {et~al.} 2009, \apj, 690, 358, \dodoi{10.1088/0004-637X/690/1/358}

\bibitem[{{Hou} {et~al.}(2009){Hou}, {Parker}, {Harris}, \& {Wilman}}]{Hou09}
{Hou}, A., {Parker}, L.~C., {Harris}, W.~E., \& {Wilman}, D.~J. 2009, \apj, 702, 1199, \dodoi{10.1088/0004-637X/702/2/1199}

\bibitem[{{Iannuzzi} \& {Dolag}(2012)}]{Iannuzzi12}
{Iannuzzi}, F., \& {Dolag}, K. 2012, \mnras, 427, 1024, \dodoi{10.1111/j.1365-2966.2012.22017.x}

\bibitem[{{Ishiyama} {et~al.}(2009){Ishiyama}, {Fukushige}, \& {Makino}}]{Ishiyama09}
{Ishiyama}, T., {Fukushige}, T., \& {Makino}, J. 2009, \pasj, 61, 1319, \dodoi{10.1093/pasj/61.6.1319}

\bibitem[{{Ishiyama} {et~al.}(2012){Ishiyama}, {Nitadori}, \& {Makino}}]{Ishiyama12}
{Ishiyama}, T., {Nitadori}, K., \& {Makino}, J. 2012, arXiv e-prints, arXiv:1211.4406.
\newblock \doarXiv{1211.4406}

\bibitem[{{Ishiyama} {et~al.}(2025){Ishiyama}, {Prada}, \& {Klypin}}]{Ishiyama2025}
{Ishiyama}, T., {Prada}, F., \& {Klypin}, A.~A. 2025, arXiv e-prints, arXiv:2503.19352, \dodoi{10.48550/arXiv.2503.19352}

\bibitem[{{Ishiyama} {et~al.}(2021){Ishiyama}, {Prada}, {Klypin}, {Sinha}, {Metcalf}, {Jullo}, {Altieri}, {Cora}, {Croton}, {de la Torre}, {Mill{\'a}n-Calero}, {Oogi}, {Ruedas}, \& {Vega-Mart{\'\i}nez}}]{Ishiyama21}
{Ishiyama}, T., {Prada}, F., {Klypin}, A.~A., {et~al.} 2021, \mnras, 506, 4210, \dodoi{10.1093/mnras/stab1755}

\bibitem[{{Klypin} {et~al.}(2016){Klypin}, {Yepes}, {Gottl{\"o}ber}, {Prada}, \& {He{\ss}}}]{Klypin16}
{Klypin}, A., {Yepes}, G., {Gottl{\"o}ber}, S., {Prada}, F., \& {He{\ss}}, S. 2016, \mnras, 457, 4340, \dodoi{10.1093/mnras/stw248}

\bibitem[{{Kravtsov} \& {Borgani}(2012)}]{Kravtsov12}
{Kravtsov}, A.~V., \& {Borgani}, S. 2012, \araa, 50, 353, \dodoi{10.1146/annurev-astro-081811-125502}

\bibitem[{{Lau} {et~al.}(2015){Lau}, {Nagai}, {Avestruz}, {Nelson}, \& {Vikhlinin}}]{Lau15}
{Lau}, E.~T., {Nagai}, D., {Avestruz}, C., {Nelson}, K., \& {Vikhlinin}, A. 2015, \apj, 806, 68, \dodoi{10.1088/0004-637X/806/1/68}

\bibitem[{{Lemze} {et~al.}(2012){Lemze}, {Wagner}, {Rephaeli}, {Sadeh}, {Norman}, {Barkana}, {Broadhurst}, {Ford}, \& {Postman}}]{Lemze12}
{Lemze}, D., {Wagner}, R., {Rephaeli}, Y., {et~al.} 2012, \apj, 752, 141, \dodoi{10.1088/0004-637X/752/2/141}

\bibitem[{{Lotz} {et~al.}(2019){Lotz}, {Remus}, {Dolag}, {Biviano}, \& {Burkert}}]{Lotz19}
{Lotz}, M., {Remus}, R.-S., {Dolag}, K., {Biviano}, A., \& {Burkert}, A. 2019, \mnras, 488, 5370, \dodoi{10.1093/mnras/stz2070}

\bibitem[{{Mamon} {et~al.}(2013){Mamon}, {Biviano}, \& {Bou{\'e}}}]{Mamon13}
{Mamon}, G.~A., {Biviano}, A., \& {Bou{\'e}}, G. 2013, \mnras, 429, 3079, \dodoi{10.1093/mnras/sts565}

\bibitem[{{Mamon} \& {Bou{\'e}}(2010)}]{Mamon10}
{Mamon}, G.~A., \& {Bou{\'e}}, G. 2010, \mnras, 401, 2433, \dodoi{10.1111/j.1365-2966.2009.15817.x}

\bibitem[{{Merritt}(1985)}]{Merritt85}
{Merritt}, D. 1985, \aj, 90, 1027, \dodoi{10.1086/113810}

\bibitem[{{Munari} {et~al.}(2013){Munari}, {Biviano}, {Borgani}, {Murante}, \& {Fabjan}}]{Munari13}
{Munari}, E., {Biviano}, A., {Borgani}, S., {Murante}, G., \& {Fabjan}, D. 2013, \mnras, 430, 2638, \dodoi{10.1093/mnras/stt049}

\bibitem[{{Navarro} {et~al.}(1996){Navarro}, {Frenk}, \& {White}}]{NFW96}
{Navarro}, J.~F., {Frenk}, C.~S., \& {White}, S.~D.~M. 1996, \apj, 462, 563, \dodoi{10.1086/177173}

\bibitem[{{Navarro} {et~al.}(1997){Navarro}, {Frenk}, \& {White}}]{NFW97}
---. 1997, \apj, 490, 493, \dodoi{10.1086/304888}

\bibitem[{{Oman} {et~al.}(2013){Oman}, {Hudson}, \& {Behroozi}}]{Oman13}
{Oman}, K.~A., {Hudson}, M.~J., \& {Behroozi}, P.~S. 2013, \mnras, 431, 2307, \dodoi{10.1093/mnras/stt328}

\bibitem[{{Planck Collaboration} {et~al.}(2016){Planck Collaboration}, {Ade}, {Aghanim}, {Arnaud}, {Ashdown}, {Aumont}, {Baccigalupi}, {Banday}, {Barreiro}, {Bartlett}, {Bartolo}, {Battaner}, {Battye}, {Benabed}, {Beno{\^\i}t}, {Benoit-L{\'e}vy}, {Bernard}, {Bersanelli}, {Bielewicz}, {Bock}, {Bonaldi}, {Bonavera}, {Bond}, {Borrill}, {Bouchet}, {Boulanger}, {Bucher}, {Burigana}, {Butler}, {Calabrese}, {Cardoso}, {Catalano}, {Challinor}, {Chamballu}, {Chary}, {Chiang}, {Chluba}, {Christensen}, {Church}, {Clements}, {Colombi}, {Colombo}, {Combet}, {Coulais}, {Crill}, {Curto}, {Cuttaia}, {Danese}, {Davies}, {Davis}, {de Bernardis}, {de Rosa}, {de Zotti}, {Delabrouille}, {D{\'e}sert}, {Di Valentino}, {Dickinson}, {Diego}, {Dolag}, {Dole}, {Donzelli}, {Dor{\'e}}, {Douspis}, {Ducout}, {Dunkley}, {Dupac}, {Efstathiou}, {Elsner}, {En{\ss}lin}, {Eriksen}, {Farhang}, {Fergusson}, {Finelli}, {Forni}, {Frailis}, {Fraisse}, {Franceschi}, {Frejsel}, {Galeotta}, {Galli}, {Ganga}, {Gauthier}, {Gerbino}, {Ghosh}, {Giard},
  {Giraud-H{\'e}raud}, {Giusarma}, {Gjerl{\o}w}, {Gonz{\'a}lez-Nuevo}, {G{\'o}rski}, {Gratton}, {Gregorio}, {Gruppuso}, {Gudmundsson}, {Hamann}, {Hansen}, {Hanson}, {Harrison}, {Helou}, {Henrot-Versill{\'e}}, {Hern{\'a}ndez-Monteagudo}, {Herranz}, {Hildebrand t}, {Hivon}, {Hobson}, {Holmes}, {Hornstrup}, {Hovest}, {Huang}, {Huffenberger}, {Hurier}, {Jaffe}, {Jaffe}, {Jones}, {Juvela}, {Keih{\"a}nen}, {Keskitalo}, {Kisner}, {Kneissl}, {Knoche}, {Knox}, {Kunz}, {Kurki-Suonio}, {Lagache}, {L{\"a}hteenm{\"a}ki}, {Lamarre}, {Lasenby}, {Lattanzi}, {Lawrence}, {Leahy}, {Leonardi}, {Lesgourgues}, {Levrier}, {Lewis}, {Liguori}, {Lilje}, {Linden-V{\o}rnle}, {L{\'o}pez-Caniego}, {Lubin}, {Mac{\'\i}as-P{\'e}rez}, {Maggio}, {Maino}, {Mandolesi}, {Mangilli}, {Marchini}, {Maris}, {Martin}, {Martinelli}, {Mart{\'\i}nez-Gonz{\'a}lez}, {Masi}, {Matarrese}, {McGehee}, {Meinhold}, {Melchiorri}, {Melin}, {Mendes}, {Mennella}, {Migliaccio}, {Millea}, {Mitra}, {Miville-Desch{\^e}nes}, {Moneti}, {Montier}, {Morgante}, {Mortlock},
  {Moss}, {Munshi}, {Murphy}, {Naselsky}, {Nati}, {Natoli}, {Netterfield}, {N{\o}rgaard-Nielsen}, {Noviello}, {Novikov}, {Novikov}, {Oxborrow}, {Paci}, {Pagano}, {Pajot}, {Paladini}, {Paoletti}, {Partridge}, {Pasian}, {Patanchon}, {Pearson}, {Perdereau}, {Perotto}, {Perrotta}, {Pettorino}, {Piacentini}, {Piat}, {Pierpaoli}, {Pietrobon}, {Plaszczynski}, {Pointecouteau}, {Polenta}, {Popa}, {Pratt}, {Pr{\'e}zeau}, {Prunet}, {Puget}, {Rachen}, {Reach}, {Rebolo}, {Reinecke}, {Remazeilles}, {Renault}, {Renzi}, {Ristorcelli}, {Rocha}, {Rosset}, {Rossetti}, {Roudier}, {Rouill{\'e} d'Orfeuil}, {Rowan-Robinson}, {Rubi{\~n}o-Mart{\'\i}n}, {Rusholme}, {Said}, {Salvatelli}, {Salvati}, {Sandri}, {Santos}, {Savelainen}, {Savini}, {Scott}, {Seiffert}, {Serra}, {Shellard}, {Spencer}, {Spinelli}, {Stolyarov}, {Stompor}, {Sudiwala}, {Sunyaev}, {Sutton}, {Suur-Uski}, {Sygnet}, {Tauber}, {Terenzi}, {Toffolatti}, {Tomasi}, {Tristram}, {Trombetti}, {Tucci}, {Tuovinen}, {T{\"u}rler}, {Umana}, {Valenziano}, {Valiviita}, {Van Tent},
  {Vielva}, {Villa}, {Wade}, {Wandelt}, {Wehus}, {White}, {White}, {Wilkinson}, {Yvon}, {Zacchei}, \& {Zonca}}]{Planck15}
{Planck Collaboration}, {Ade}, P.~A.~R., {Aghanim}, N., {et~al.} 2016, \aap, 594, A13, \dodoi{10.1051/0004-6361/201525830}

\bibitem[{{Praton} \& {Schneider}(1994)}]{Praton94}
{Praton}, E.~A., \& {Schneider}, S.~E. 1994, \apj, 422, 46, \dodoi{10.1086/173702}

\bibitem[{{The} \& {White}(1986)}]{The86}
{The}, L.~S., \& {White}, S.~D.~M. 1986, \aj, 92, 1248, \dodoi{10.1086/114258}

\bibitem[{{Tiret} {et~al.}(2007){Tiret}, {Combes}, {Angus}, {Famaey}, \& {Zhao}}]{Tiret07}
{Tiret}, O., {Combes}, F., {Angus}, G.~W., {Famaey}, B., \& {Zhao}, H.~S. 2007, \aap, 476, L1, \dodoi{10.1051/0004-6361:20078569}

\bibitem[{{Tyler} {et~al.}(2013){Tyler}, {Rieke}, \& {Bai}}]{Tyler13}
{Tyler}, K.~D., {Rieke}, G.~H., \& {Bai}, L. 2013, \apj, 773, 86, \dodoi{10.1088/0004-637X/773/2/86}

\bibitem[{{Vikhlinin} {et~al.}(2009){Vikhlinin}, {Burenin}, {Ebeling}, {Forman}, {Hornstrup}, {Jones}, {Kravtsov}, {Murray}, {Nagai}, {Quintana}, \& {Voevodkin}}]{Vikhlinin09}
{Vikhlinin}, A., {Burenin}, R.~A., {Ebeling}, H., {et~al.} 2009, \apj, 692, 1033, \dodoi{10.1088/0004-637X/692/2/1033}

\bibitem[{{Voit}(2005)}]{Voit05}
{Voit}, G.~M. 2005, Reviews of Modern Physics, 77, 207, \dodoi{10.1103/RevModPhys.77.207}

\bibitem[{{Wetzel}(2011)}]{Wetzel11}
{Wetzel}, A.~R. 2011, \mnras, 412, 49, \dodoi{10.1111/j.1365-2966.2010.17877.x}

\bibitem[{{Wetzel} \& {Nagai}(2015)}]{Wetzel15}
{Wetzel}, A.~R., \& {Nagai}, D. 2015, \apj, 808, 40, \dodoi{10.1088/0004-637X/808/1/40}

\bibitem[{{Wojtak} \& {{\L}okas}(2010)}]{Wojtak10}
{Wojtak}, R., \& {{\L}okas}, E.~L. 2010, \mnras, 408, 2442, \dodoi{10.1111/j.1365-2966.2010.17297.x}

\bibitem[{{Wojtak} {et~al.}(2009){Wojtak}, {{\L}okas}, {Mamon}, \& {Gottl{\"o}ber}}]{Wojtak09}
{Wojtak}, R., {{\L}okas}, E.~L., {Mamon}, G.~A., \& {Gottl{\"o}ber}, S. 2009, \mnras, 399, 812, \dodoi{10.1111/j.1365-2966.2009.15312.x}

\end{thebibliography}
\bibliographystyle{aasjournal}

\appendix

\end{document}